\documentclass{aa}

\usepackage{longtable}
\usepackage{graphicx}
\usepackage{txfonts}
\usepackage{natbib}
\bibpunct{(}{)}{;}{a}{}{,} 

\begin{document}

\title{A dynamical study of Galactic globular clusters \\ under different relaxation conditions}
\subtitle{}
\author{A. Zocchi \and G. Bertin \and A. L. Varri}
\institute{Universit\`a degli Studi di Milano, Dipartimento di Fisica, via Celoria 16, I-20133 Milano, Italy \\
           \email{alice.zocchi@unimi.it}}
\date{Received XXX; accepted YYY}

\abstract
   {}
{We perform a systematic combined photometric and kinematic analysis of a sample of globular clusters under different relaxation conditions, based on their core relaxation time (as listed in available catalogs), by means of two well-known families of spherical stellar dynamical models. Systems characterized by shorter relaxation time scales are expected to be better described by isotropic King models, while less relaxed systems might be interpreted by means of non-truncated, radially-biased anisotropic $f^{(\nu)}$ models, originally designed to represent stellar systems produced by a violent relaxation formation process and applied here for the first time to the study of globular clusters.}
{The comparison between dynamical models and observations is performed by fitting simultaneously surface brightness and velocity dispersion profiles. For each globular cluster, the best-fit model in each family is identified, along with a full error analysis on the relevant parameters. Detailed structural properties and mass-to-light ratios are also explicitly derived.}
{We find that King models usually offer a good representation of the observed photometric profiles, but often lead to less satisfactory fits to the kinematic profiles, independently of the relaxation condition of the systems. For some less relaxed clusters, $f^{(\nu)}$ models provide a good description of both observed profiles. Some derived structural characteristics, such as the total mass or the half-mass radius, turn out to be significantly model-dependent. The analysis confirms that, to answer some important dynamical questions that bear on the formation and evolution of globular clusters, it would be highly desirable to acquire larger numbers of accurate kinematic data-points, well distributed over the cluster field.}
   {}


\maketitle


\section{Introduction}
\label{Sect.1}

It is well known that for most globular clusters the relevant relaxation times are shorter than their age. Therefore, since the two-body relaxation processes have had time to act, these systems may be considered to be close to thermodynamical relaxation, with a distribution function that should be close to a Maxwellian.

The spherical King models \citep{King1966} are defined by a distribution function of this type, with a truncation in energy introduced to mimic the role of tides (for the relevant definition and a summary of the main properties, see App.~\ref{App.A.1}). In general, they are thought to be a physically justified simple tool to represent quasi-relaxed stellar systems, in particular well suited to describe globular clusters. In this context, King models are usually considered as the correct zeroth-order dynamical reference model. The available catalogs (e.g., see \citealp{HarrisCat2010}) record a number of structural properties of globular clusters that are derived from an interpretation of observational data based on the use of King models. The family of King models is also frequently used in dynamical studies aimed at identifying given physical mechanisms and represents a most popular choice of initial conditions for numerical simulations designed to investigate the dynamical evolution of star clusters (e.g., see \citealp{ChernoffWeinberg1990}, and, more recently, \citealp{GielesBaumgardt2008}).

In reality, this simple physical picture suffers from a number of limitations, which become more evident now that much improved observations allow us to go well beyond a zeroth-order picture of globular clusters. Of course, numerical simulations can now model in a realistic way the important effects of discreteness, mass segregation, binaries, rotation, core collapse, and time-dependent tides (e.g., see \citealp{Kuepper2010}), which are known to govern the evolution of globular clusters. Still, in relation to the general issue of relaxation it would be interesting to test further to what extent King models (or alternative idealized models) actually work as a useful paradigm to represent small, quasi-relaxed stellar systems.

From the observational point of view, there remains indeed one unsatisfactory point, which this paper would like to address, at least on the basis of the currently available data. This is that the application (and thus the test) of the spherical King models is frequently carried out, in practice, only in relation to the available photometric profiles (see \citealp{TKD1995} and \citealp{MLvdM2005}), without the corresponding tests on the associated kinematical profiles, in contrast to what is routinely done in the studies of early-type galaxies. Some studies of Galactic globular clusters, constrained simultaneously by density and velocity dispersion profiles, are actually available, mainly based on multi-mass Michie-King models (e.g., see \citealp{GunnGriffin1979}, \citealp{Meylan1995}, and \citealp{MeylanMayor1991} for M3, $\omega$ Centauri, and NGC $6397$, respectively), but a systematic and homogeneous investigation is still missing. Note that, on the galactic side, deep investigations of these issues, starting with the mid 1970s, have led to the remarkable discovery that bright ellipticals are generally supported by anisotropic pressure and contain significant amounts of dark matter inside the effective radius $R_\mathrm{e}$ (e.g., see Chap.~24 in \citealp{libroverde}). From the study of elliptical galaxies it has also been learned that structurally different models (as diagnosed by their kinematics or characterized by their virial coefficients) may have remarkably similar photometric profiles (e.g., see Appendix~D of \citealp{BertinCiottiDelPrincipe2002}).

In this paper, \textit{we wish to check whether indeed King models perform better in more relaxed systems}, by studying the combined photometric and kinematical profiles for a sample of globular clusters. We will refer to the surface brightness profiles collected by \citet{TKD1995} (supplemented by more recent data, when available), in order to deal with a homogeneous sample, and limit our discussion to one-component dynamical models (leaving aside the issues of core collapse and mass segregation; we will thus exclude from our sample clusters with evidence of core collapse). Our sample will then be basically defined by the requirement that a sufficient number of kinematical data-points is available from the literature, so as to define a reasonably accurate and radially extended kinematic profile for a test of a dynamical model at the global level.

Many large globular clusters have very long relaxation times. Therefore, we have decided to model the same data also by means of models (the $f^{(\nu)}$ models; see \citealp{BertinTrenti2003}) explicitly constructed for the context of violently relaxed elliptical galaxies. In other words, \textit{we wish to check whether less relaxed clusters tend to conform to the picture of formation via incomplete violent relaxation}, which has the characteristic signature of radially-biased pressure anisotropy for less bound stars. The use of the $f^{(\nu)}$ models is preferred to other options (for example to the use of King-Michie models, \citealp{Michie1963}), because these models are based on a detailed physical justification and have been shown to perform well both in relation to the observations of bright ellipticals and to the properties of the products of incomplete violent relaxation found in numerical simulations of collisionless collapse (over a range of nine orders of magnitude in the computed density profiles, with an excellent fit to the properties of the generated pressure anisotropy profiles; see Trenti, Bertin, van Albada 2005). For a fair comparison with the King models we should have referred to \textit{truncated} $f^{(\nu)}$ models; for the present simple exploratory investigation, and in order to keep the comparison between models characterized by the same number of parameters, we have decided to use the non-truncated $f^{(\nu)}$ models.

The paper is organized as follows. In Sect.~\ref{Sect.2} we introduce the sample of globular clusters selected for this study. In Sect.~\ref{Sect.3} we describe the available data sets and the procedure we followed to obtain the radial profiles used in our analysis. In Sect.~\ref{Sect.4} we show the results of our work and in Sect.~\ref{Sect.5} we draw our conclusions. A brief description of the characteristics of the dynamical models is provided in Appendix~\ref{App.A}; in Appendix~\ref{App.B} we describe the adopted fitting procedure.


\section{The selected sample}
\label{Sect.2}

In this paper we wish to analyze globular clusters characterized by different relaxation conditions, measured by the central (core) relaxation time $T_{\mathrm{c}}$ as listed in the Harris catalog \citep{HarrisCat2010}. We decided to order globular clusters according to the core relaxation time, rather than to the half-mass relaxation time, because in general $T_{\mathrm{c}}$ is less model-dependent (as pointed out in Sect.~\ref{Subsect.4.2}). We checked that, by ordering the sample with respect to the half-mass relaxation time, some changes in the composition of the three classes identified below would occur (this fact can be easily seen by inspecting the values of these time scales for the globular clusters listed in Table~\ref{table1}), but with minor effects in relation to the general conclusions of the paper.

Looking at the values of the relevant relaxation times recorded in the available catalog for the globular cluster system of our Galaxy, we see that the distinction between relaxed and partially relaxed globular clusters is not sharp. Therefore, we introduce a simple criterion to classify globular clusters according to their relaxation state. We consider three relaxation classes: relaxed globular clusters, for which $\log{T_{\mathrm{c}}}<8$ (first class); globular clusters in an intermediate relaxation condition, for which $8<\log{T_{\mathrm{c}}}<9$ (second class); partially relaxed globular clusters, for which $\log{T_{\mathrm{c}}}>9$ (third class; $T_{\mathrm{c}}$ is expressed in years).

The sample of globular clusters has been selected on the basis of the following criteria: (i) We exclude post-core-collapse globular clusters, that is, we reject the clusters labeled as post-core-collapse by \citet{HarrisCat2010}. The reason is that we wish to test King models on the global scale, avoiding the subtle modeling issues that characterize the central regions of these systems, especially if phenomena a priori known to go beyond the King modeling are involved. (ii) We choose globular clusters for which an accurate and extended surface brightness profile is available. (iii) We select clusters for which at least $140$ stellar radial (line-of-sight) velocities have already been measured. We impose a lower limit to the number of measured velocities, because we wish to extract from such data a reasonably well-defined velocity dispersion profile; the value of this limit is fixed in a way that allows us to include in the analysis globular clusters that belong to the different relaxation classes defined previously. (iv) We exclude from our list all the clusters that have less than $35$ stellar radial velocities inside the projected half-light radius $R_\mathrm{e}$ (also called effective radius; the values of this quantity are reported in the Harris catalog, where the notation is $r_\mathrm{h}$). We introduce this further requirement, because we wish to analyze velocity dispersion profiles that characterize the stellar systems on the largest radial extent. For bright elliptical galaxies the kinematical data-points inside $R_\mathrm{e}$ are usually the easiest to get, and often turn out to discriminate among very different models. To test how well the King models perform, the central regions are a natural ground for comparison with other models.

\begin{table*}
\caption[]{The selected globular clusters.}
\label{table1}
\centering
\begin{tabular}{l cccc cccccc c}
\hline\hline
\multicolumn{1}{c}{Globular Cluster} & $d_{\sun}$ & $C$ & $\log{T_{\mathrm{c}}}$ & $\log{T_\mathrm{M}}$ & $e$ & $v_{\mathrm{rot}}$ & $N_{\mathrm{v}}$ & $N_{\mathrm{e}}$ & $\frac{R_{\mathrm{K}}}{R_{\mathrm{e}}}$ & $\frac{R_{\mathrm{K}}}{r_{\mathrm{tr}}}$ & Ref. \\
\hline
NGC  362                &  8.6 & 1.76 & 7.76 &  8.93 & 0.01 & 0.0  &  208 &   92 &  4.22 & 0.33 & (1)      \\
NGC 7078 (M15)          & 10.4 & 2.29 & 7.84 &  9.32 & 0.05 & 1.7  & 1777 & 1298 & 16.94 & 0.62 & (2)      \\
NGC  104 (47 Tuc)       &  4.5 & 2.07 & 7.84 &  9.55 & 0.09 & 2.2  & 2638 &  709 & 19.27 & 1.44 & (3), (4) \\
NGC 6121 (M4)           &  2.2 & 1.65 & 7.90 &  8.93 & 0.00 & 0.9  &  200 &   55 & 10.36 & 0.87 & (4)      \\
NGC 6341 (M92)          &  8.3 & 1.68 & 7.96 &  9.02 & 0.10 & 2.5  &  295 &   42 & 13.96 & 1.14 & (5)      \\
\hline
NGC 6218 (M12)          &  4.8 & 1.34 & 8.19 &  8.87 & 0.04 & 0.15 &  242 &   58 & 10.38 & 1.06 & (4)      \\
NGC 6254 (M10)          &  4.4 & 1.38 & 8.21 &  8.90 & 0.00 & $\ldots$ &  147 &   47 &  5.22 & 0.55 & (6)      \\
NGC 6656 (M22)          &  3.2 & 1.38 & 8.53 &  9.23 & 0.14 & 1.5  &  345 &  116 &  8.40 & 0.88 & (4)      \\
NGC 3201                &  4.9 & 1.29 & 8.61 &  9.27 & 0.12 & 1.2  &  399 &  201 & 10.35 & 1.27 & (7)      \\
NGC 6809 (M55)          &  5.4 & 0.93 & 8.90 &  9.29 & 0.02 & 0.25 &  728 &  311 &  7.79 & 1.44 & (4)      \\
NGC  288                &  8.9 & 0.99 & 8.99 &  9.32 & $\ldots$ & 0.25 &  171 &   68 &  5.53 & 0.93 & (4), (6) \\
\hline
NGC 5139 ($\omega$ Cen) &  5.2 & 1.31 & 9.60 & 10.09 & 0.17 & 7.9  & 2060 &  554 &  5.97 & 0.62 & (8), (9)      \\
NGC 2419                & 82.6 & 1.37 & 9.87 & 10.63 & 0.03 & 0.6  &  166 &   38 & 14.63 & 1.74 & (10)      \\
\hline
\end{tabular}
\tablefoot{From left to right, the following quantities are displayed: distance from the Sun (kpc), concentration parameter, logarithm of the core relaxation time (years), logarithm of the half-mass relaxation time (years), ellipticity, rotational velocity (km~s$^{-1}$), total number of velocity data-points available, number of velocity data-points inside the projected half-light radius, ratio of the radius of the outermost velocity point to the projected half-light radius, and ratio of the radius of the outermost velocity point to the truncation radius. The sources of the kinematical data are listed in the last column (see main text for references of the other quantities).}
\tablebib{(1) \citealp{Fischer1993}; (2) \citealp{Gebhardt2000}; (3) \citealp{Gebhardt1995}; (4) \citealp{Lane2011}; (5) \citealp{Drukier2007}; (6) \citealp{Carretta2009}; (7) \citealp{Cote1995}; (8) \citealp{Mayor1997}; (9) \citealp{Reijns2006}; (10) \citealp{Ibata2011}.}
\end{table*}

The most restrictive elements in identifying a significant sample of globular clusters are the requirements on the radial velocity data, because for only few globular clusters the desired data are available. Indeed, of the $28$ Galactic globular clusters with a reasonable number of radial velocities (i.e., at least $40$ line-of-sight velocity measures on the entire spatial extent of the cluster), $3$ are flagged as post-core-collapse, $9$ have less than $140$ velocity data, and $3$ have less than $35$ data inside their projected half-light radius. In the Harris catalog, NGC $362$ and NGC $7078$ are indicated as possible post-core-collapse clusters, but we decided to keep them because, according to their concentration parameter in the Harris catalog, it is still possible to obtain an acceptable fit with King models. In this way, we are left with $13$ globular clusters that match our selection criteria. In relation to the relaxation classes defined above, our set of globular clusters contains $5$ well-relaxed clusters, $6$ clusters in an intermediate relaxation condition, and $2$ partially relaxed clusters.

To better characterize our sample in terms of the radial extent of their radial velocity data, we consider the ratio of the radius of the last kinematical point to the projected half-light radius\footnote{We always indicate with $R$ the projected (two-dimensional) radial scales, and with $r$ the intrinsic (three-dimensional) radial scales.}, $R_{\mathrm{K}}/R_{\mathrm{e}}$, and the ratio of the radius of the last kinematical point to the truncation radius\footnote{In the latest version of the Harris catalog \citep{HarrisCat2010}, the values of the truncation radii are not listed; we calculated them from the available values of $R_{\mathrm{c}}$ and $C$ (see Appendix~\ref{App.A.1}), as: $r_{\mathrm{tr}} = R_{\mathrm{c}}\,10^{C}$ (see notes in the Harris catalog bibliography). For the three core-collapsed globular clusters in Table~\ref{table2}, for which the concentration parameter is arbitrarily set to $C = 2.50$, the value of $r_{\mathrm{tr}}$ should be considered only as indicative.}, $R_{\mathrm{K}}/r_{\mathrm{tr}}$. We judge the following values of the two ratios, $R_{\mathrm{K}}/R_{\mathrm{e}}\geq3$ and $R_{\mathrm{K}}/r_{\mathrm{tr}}\geq0.8$, to be satisfactory. All the selected globular clusters satisfy the first relation, and all but four globular clusters satisfy the second condition.

Table~\ref{table1} gives the sample of selected globular clusters, listed in order of increasing core relaxation time $\log{T_{\mathrm{c}}}$. The first part of the table contains relaxed globular clusters, the second part those in an intermediate relaxation condition, and partially relaxed clusters are shown in the last part. For each object, the values of the adopted cluster distance from the Sun $d_{\sun}$ (expressed in kpc), the concentration parameter $C$, the logarithm of the core relaxation time $\log{T_{\mathrm{c}}}$, the logarithm of the half-mass relaxation time $\log{T_{\mathrm{M}}}$ (where $T_{\mathrm{c}}$ and $T_{\mathrm{M}}$, in other papers often indicated with the symbols $t_{\mathrm{rc}}$ and $t_{\mathrm{rh}}$, are expressed in years) and the ellipticity $e$ are recorded (as listed in the Harris 2010 catalog). In addition, the maximum rotational velocity $v_{\mathrm{rot}}$ (in km~s$^{-1}$; the references for these values are \citealp{Lane2011} for the globular clusters for which we use the kinematic data published in this paper, and \citealp{MeylanHeggie1997} for the others), the number of velocity data-points available $N_{\mathrm{v}}$, the number of velocity data-points inside the projected half-light radius $N_{\mathrm{e}}$, the ratios of the radius of the outermost velocity point $R_{\mathrm{K}}$ to the projected half-light radius $R_{\mathrm{e}}$ (in the Harris catalog and in other papers often indicated as $r_{\mathrm{h}}$) and to the truncation radius $r_{\mathrm{tr}}$, and the sources of the kinematic data are given in the last columns.

For completeness, in Table~\ref{table2} we list globular clusters with a fairly rich kinematical information available (line-of-sight velocity measurements for at least 40 stars), but rejected in the present analysis because they do not satisfy our selection criteria. The table is organized in three parts, thus showing which criterion is not met: in the first part we list post-core-collapse globular clusters, in the second part those with an insufficient total number of kinematic data ($N_{\mathrm{v}}$), and in the third part those with an insufficient number of data inside the half-light radius ($N_{\mathrm{e}}$). In each part of the table, globular clusters are listed in order of increasing number of total stellar velocity data $N_{\mathrm{v}}$. Column entries are the same as in Table~\ref{table1} (except for cluster distance, ellipticity, and rotational velocity, which are not recorded).


\section{The data sets}
\label{Sect.3}

\subsection{The surface brightness profiles}
\label{Subsect.3.1}

To deal with a homogeneous sample, we decided to use the surface brightness profiles provided by \citet{TKD1995} (this is the same starting point of \citealp{MLvdM2005}). This choice guarantees that the profiles have been constructed with the same method, even though the actual data come from different sources.

\begin{table*}
\caption[]{Rejected globular clusters. For a description of column entries, see Table~\ref{table1}.}
\label{table2}
\centering
\begin{tabular}{lcccccccc}
\hline\hline
\multicolumn{1}{c}{Globular Cluster} & $C$ & $ \log{T_{\mathrm{c}}}$ & $\log{T_{\mathrm{M}}}$ & $N_{\mathrm{v}}$ & $N_{\mathrm{e}}$ & $\frac{R_{\mathrm{K}}}{R_{\mathrm{e}}}$ & $\frac{R_{\mathrm{K}}}{r_{\mathrm{tr}}}$ & Ref. \\
\hline
NGC 6397        & 2.50 & 4.94 &  8.60 & 103 & 103 &  0.37 & 0.07 & (1) \\
NGC 7099 (M30)  & 2.50 & 6.37 &  8.88 & 196 &  33 & 11.20 & 0.61 & (2) \\
NGC 6752        & 2.50 & 6.88 &  8.87 & 437 &  75 & 15.24 & 0.54 & (2) \\
\hline
IC  4499        & 1.21 & 9.21 &  9.73 &  43 &  11 &  6.93 & 0.87 & (3) \\
NGC 6171 (M107) & 1.53 & 8.06 &  9.00 &  66 &  37 &  3.67 & 0.40 & (4) \\
NGC 5466        & 1.04 & 9.35 &  9.76 &  66 &  49 &  1.69 & 0.20 & (5) \\
NGC 7089 (M2)   & 1.59 & 8.48 &  9.40 &  69 &  19 &  8.02 & 0.68 & (6) \\
NGC 5053        & 0.74 & 9.81 &  9.87 &  71 &  27 &  3.28 & 0.75 & (7) \\
NGC 4590 (M68)  & 1.41 & 8.45 &  9.27 & 122 &  29 &  6.69 & 0.68 & (8) \\
NGC 2808        & 1.56 & 8.24 &  9.15 & 123 &   7 &  8.41 & 0.74 & (9) \\
NGC 6205 (M13)  & 1.53 & 8.51 &  9.30 & 123 &  10 &  7.92 & 0.64 & (10) \\
NGC 5904 (M5)   & 1.73 & 8.28 &  9.41 & 136 &  18 &  6.61 & 0.50 & (8) \\
\hline
NGC 1904 (M79)  & 1.70 & 7.83 &  8.95 & 146 &   2 & 14.46 & 1.17 & (11) \\
NGC 5024 (M53)  & 1.72 & 8.73 &  9.76 & 180 &   4 & 16.24 & 1.16 & (2) \\
NGC 1851        & 1.86 & 7.43 &  8.82 & 184 &   0 & 21.21 & 1.66 & (11) \\
\hline
\end{tabular}
\tablebib{(1) \citealp{Gebhardt1995}; (2) \citealp{Lane2011}; (3) \citealp{Hankey2010}; (4) \citealp{Piatek1994}; (5) \citealp{Shetrone2010}; (6) \citealp{Pryor1986}; (7) \citealp{Yan1996}; (8) \citealp{Carretta2009}; (9) \citealp{Carretta2006}; (10) \citealp{Meszaros2009}; (11) \citealp{Scarpa2011}.}
\end{table*}

For each globular cluster, the profile is composed of $N_{\mathrm{p}}$ photometric data-points, given by $\log R_i$, the logarithm of the radius $R_i$, measured in arcsec, and by $m_{\mathrm{V}}(R_i)$, the V band surface brightness measured in mag~arcsec$^{-2}$ at the radial position $R_i$. The data also include $m_{\mathrm{V,C}}(R_i)$, the surface brightness calculated with the Chebyshev polynomials, which provides an accurate approximation to the overall profile; $m_{\mathrm{V}}(R_i)-m_{\mathrm{V,C}}(R_i)$, the Chebyshev residual; $w_i$, a weight that the authors assign to each measurement. All the surface brightness profiles considered in this paper are relative to the V band.

These data have to be properly corrected and treated before a comparison can be made with the theoretical models. First, we introduced an extinction correction, under the assumption that such extinction can be considered to be constant over the entire extent of each globular cluster. We calculated the extinction $A_{\mathrm{V}}$ from the reddening listed in the Harris catalog and obtained $m(R_i) = m_{\mathrm{V}}(R_i)-A_{\mathrm{V}}$, for $i = 1, \ldots, N_{\mathrm{p}}$. This is the only correction applied to the data: we assume that \citet{TKD1995} already removed any foreground and background contamination that could affect the measurements. Then, we followed the procedure described by \citet{MLvdM2005} to estimate the uncertainties $\delta m_i$ on the data: starting from the weights, we calculated\footnote{$\sigma_{\mu}$ is a constant that varies from cluster to cluster, the value of which can be found in Table~$6$ in \citet{MLvdM2005}.} $\delta m_i = \sigma_{\mu}/w_i$ (note that in the fits we use only the points with weights $w_i \geq 0.15$ in the original profile, as suggested by \citealp{MLvdM2005}).

Usually, the less reliable parts in the profiles published by \citet{TKD1995} are the central regions. Therefore, we decided to combine these profile with the more recent and accurate surface brightness profiles by \citet{Noyola2006}, when available. For the globular clusters NGC 104 and NGC 6341, we simply combined the profiles from the two sources; for NGC 6254 and NGC 7078, we decided to combine the two data sets by removing the points from \citet{TKD1995}, when they do not match the profile by \citet{Noyola2006} (the removed points are inside $3.5 \arcsec$ and $14.5 \arcsec$, respectively). In the case of NGC 7078, it should be emphasized that, with this treatment, the profile changes significantly in the central regions, and the central slope becomes steeper.

In the case of NGC $5139$ ($\omega$ Cen) we decided to add to the surface brightness profile the inner points that Eva Noyola kindly provided us \citep{Noyola2008}; still the number of data-points in our final composite surface brightness profile of this cluster is significantly smaller than that of the other clusters.

\subsection{The velocity dispersion profiles}
\label{Subsect.3.2}

We divided the data (see Table~\ref{table1} for detailed references) in several radial bins containing an equal number of stars; we chose the binning that represents the best compromise between having a rich profile and having accurate points (by increasing the number of bins, that is by decreasing the number of points per bin, the errors on the velocity dispersion increase). In principle, we might consider using unbinned data, to avoid loss of information (for the study of dwarf spheroidals, see \citealp{Wilkinson2002}).  Here we preferred to follow the more traditional approach of constructing the associated one-dimensional profiles, a method that can be applied in a similar way to both kinematical and photometric data and that allows us to follow well-established fitting procedures used in the past (especially in studies of elliptical galaxies). For the clusters with less numerous kinematical data-points, we did experiment with changing the radial binning and checked the related consequences.

The method used to calculate the mean velocity and the velocity dispersion from stellar radial velocities is basically the one described by \citet{PryorMeylan1993}. We started by calculating the mean $v_{\mathrm{r}}$ and the dispersion $\sigma_{\mathrm{c}}$ for the entire set of velocities. The mean velocity represents the overall velocity of the entire cluster. Then, taking this value as a constant for the entire cluster, we calculated the line-of-sight velocity dispersion $\sigma(R_i)$ and the related accuracy $\delta \sigma_i$ inside the bins in which the data have been divided. For each bin, we indicate the distance from the center $R_i$ as the mean of the radial positions of the stars that it contains. In this paper we ignore the possible presence of rotation and therefore consider the various kinematical data-points in each bin, after subtraction of the systemic velocity, to contribute only to velocity dispersion (random motions).

\begin{table*}[t]
\caption[]{Dimensionless parameters and physical scales of the best-fit models.}
\label{table3}
\centering
\begin{tabular}{rcccccccccc}
\hline\hline
 & & & \multicolumn{2}{c}{King Models} & & & & \multicolumn{2}{c}{$f^{(\nu)}$ Models} & \\
\cline{3-6}  \cline{8-11}
NGC  & & $\Psi$ & $r_{\mathrm{0}}$ & $\mu_{\mathrm{0}}$ & $V$ & & $\Psi$ & $r_{\mathrm{scale}}$ & $\mu_{\mathrm{0}}$ & $V$ \\
 (1) & & (2) & (3) & (4) & (5) & & (6) & (7) & (8) & (9) \\
\hline
104  & & 8.58 $\pm$ 0.01 &  23.09 $\pm$ 0.23 & 14.33 $\pm$ 0.01 & 12.27 $\pm$ 0.19  & & 8.21 $\pm$ 0.02 & 250.44 $\pm$  1.46 & 14.29 $\pm$ 0.01 & 14.07 $\pm$ 0.22 \\
288  & & 4.82 $\pm$ 0.10 &  91.03 $\pm$ 2.86 & 20.02 $\pm$ 0.03 &  2.85 $\pm$ 0.19  & & 3.91 $\pm$ 0.24 &  79.54 $\pm$  5.37 & 19.88 $\pm$ 0.03 &  3.70 $\pm$ 0.25 \\
362  & & 7.80 $\pm$ 0.03 &  10.21 $\pm$ 0.11 & 14.66 $\pm$ 0.01 &  8.31 $\pm$ 0.44  & & 6.86 $\pm$ 0.05 &  52.73 $\pm$  1.09 & 14.70 $\pm$ 0.01 &  9.26 $\pm$ 0.50 \\
2419 & & 6.62 $\pm$ 0.04 &  19.70 $\pm$ 0.31 & 19.43 $\pm$ 0.03 &  5.04 $\pm$ 0.43  & & 4.24 $\pm$ 0.09 &  29.86 $\pm$  0.83 & 19.44 $\pm$ 0.04 &  7.28 $\pm$
0.61 \\
3201 & & 6.17 $\pm$ 0.11 &  76.99 $\pm$ 3.05 & 18.35 $\pm$ 0.08 &  4.28 $\pm$ 0.19  & & 4.09 $\pm$ 0.40 & 103.48 $\pm$ 11.71 & 18.33 $\pm$ 0.08 &  5.01 $\pm$
0.23 \\
5139 & & 6.27 $\pm$ 0.05 & 136.94 $\pm$ 2.33 & 16.42 $\pm$ 0.04 & 14.83 $\pm$ 0.25  & & 4.31 $\pm$ 0.07 & 150.63 $\pm$  3.07 & 16.35 $\pm$ 0.04 & 23.41 $\pm$
0.40 \\
6121 & & 7.32 $\pm$ 0.07 &  74.56 $\pm$ 1.76 & 17.00 $\pm$ 0.11 &  4.01 $\pm$ 0.30  & & 7.39 $\pm$ 0.09 & 464.49 $\pm$ 20.40 & 17.01 $\pm$ 0.11 &  4.21 $\pm$
0.31 \\
6218 & & 6.11 $\pm$ 0.07 &  51.56 $\pm$ 1.51 & 17.65 $\pm$ 0.07 &  3.93 $\pm$ 0.30  & & 4.00 $\pm$ 0.14 &  60.69 $\pm$  2.52 & 17.57 $\pm$ 0.07 &  5.58 $\pm$
0.42 \\
6254 & & 6.26 $\pm$ 0.04 &  53.63 $\pm$ 0.60 & 16.88 $\pm$ 0.09 &  6.21 $\pm$ 0.37  & & 2.67 $\pm$ 0.13 &  51.48 $\pm$  1.45 & 16.84 $\pm$ 0.09 &  9.69 $\pm$
0.59 \\
6341 & & 7.54 $\pm$ 0.02 &  14.72 $\pm$ 0.13 & 15.31 $\pm$ 0.01 &  9.28 $\pm$ 0.41  & & 5.99 $\pm$ 0.04 &  50.00 $\pm$   0.80 & 15.44 $\pm$ 0.01 & 12.77 $\pm$
0.56 \\
6656 & & 6.47 $\pm$ 0.11 &  86.18 $\pm$ 2.34 & 16.41 $\pm$ 0.11 &  6.47 $\pm$ 0.38  & & 5.99 $\pm$ 0.26 & 241.58 $\pm$ 26.46 & 16.41 $\pm$ 0.11 &  7.19 $\pm$
0.42 \\
6809 & & 4.44 $\pm$ 0.11 & 129.11 $\pm$ 4.06 & 19.12 $\pm$ 0.04 &  2.92 $\pm$ 0.13  & & 3.92 $\pm$ 0.19 & 101.53 $\pm$  5.39 & 18.99 $\pm$ 0.04 &  3.61 $\pm$
0.17 \\
7078 & & 8.09 $\pm$ 0.02 &   7.72 $\pm$ 0.13 & 14.07 $\pm$ 0.03 & 11.83 $\pm$ 0.24  & & 8.17 $\pm$ 0.05 &  65.88 $\pm$  0.94 & 13.59 $\pm$ 0.06 & 12.79 $\pm$
0.26 \\
\hline
\end{tabular}
\tablefoot{For each cluster, named in column (1), for King and $f^{(\nu)}$ models, we list: the concentration parameter $\Psi$, in other papers often indicated as $W_0$ (Col. (2) and (6)), the scale radius, expressed in arcsec ($r_0$ in Col.~(3) and $r_{\mathrm{scale}}$ in Col.~(7), as defined in equations (\ref{r0}) and (\ref{scalesfnu}); note that they are intrinsic quantities; they are recorded here in arcseconds, for easier comparison with the observations, as shown in Figs.~\ref{figure1}-\ref{figure3}), the V band central surface brightness $\mu_{\mathrm{0}}$ in mag~arcsec$^{-2}$ (Col.~(4) and (8)), and the central line-of-sight velocity dispersion $V$ in km~s$^{-1}$ (Col.~(5) and (9)). Formal errors on the various parameters are also recorded (see Appendix~\ref{App.B.3}).}
\end{table*}

For the majority of the clusters in our sample, only one data-set of stellar radial velocities is available; in the following we discuss in detail the cases in which a composition of different data-sets has been performed or which require some additional comments. For NGC 104, two data-sets of radial velocities are available; we noticed that the data from \citet{Gebhardt1995} are more centrally concentrated than those from \citet{Lane2011}.\footnote{\citet{GierszHeggie2011} make some cautionary remarks about the velocity dispersion profile reported by these authors. In particular, the selection criteria adopted by \citet{Lane2011} could lead to the exclusion of some high-velocity stars, with consequent lowering of the central velocity dispersion, and to the inclusion of nonmember stars affecting the outer part of the profile. By using the composite data-set described above, we should be able to obtain reliable values of the velocity dispersion in the central regions, while the outermost points may still be affected by the inclusion of nonmember stars.} In order to have a complete sampling on the entire radial extent of the cluster, we decided to define a mixed data-set, composed of 499 data from \citet{Gebhardt1995}, located inside $100 \arcsec$, and 2139 data from \citet{Lane2011}, located outside that radius. The second case is cluster NGC 288, studied by \citet{Carretta2009} and by \citet{Lane2011}. Since these papers publish the coordinates of each star, we were able to single out the stars in common between the two data-sets: for the stars in the overlap, velocity measures by \citet{Carretta2009}, being more accurate, have been preferred. Finally, we excluded three stars for which the value of the velocity deviates by more than $4 \sigma_{\mathrm{c}}$ from the mean radial velocity $v_{\mathrm{r}}$, obtaining a final sample of 171 data.

For the globular clusters NGC 6121, NGC 6656 and NGC 6809, in addition to recent data from \citet{Lane2011}, older radial velocity measures are available in the literature, but we decided to consider only the data-sets from \citet{Lane2011}, because they are more complete and more radially extended.

In the case of NGC 5139, we merged the largest sample of velocity data available \citep{Reijns2006} with the sample provided by \citet{Mayor1997}, which provides measurements for stars located in the central region of the cluster. A delicate issue regarding this cluster is the controversial position of its center, which plays an important role also in our analysis, because we wish to build a radial-dependent velocity dispersion profile, starting from stellar positions expressed in right ascension and declination (for \citealp{Reijns2006}). To carry out a proper merging of the two data-sets, we have used the position of the center proposed by \citet{Mayor1997}; to calculate radial distances, we have followed the procedure described by \citet{vandeVen2006}.


\section{Results}
\label{Sect.4}

\subsection{Relaxation classes}
\label{Subsect.4.1}

The values of the dimensionless parameters and the physical scales of the two families of models determined by the photometric and kinematic fits are presented in Table~\ref{table3} (the fitting procedure is described in Appendix~\ref{App.B}). Note that, in general, for the $f^{(\nu)}$ models relatively low values of the concentration parameter $\Psi$ are identified, consistent with the comment given at the end of Appendix~\ref{App.A.2}. Quantitative information about the best-fit models and the observational profiles, such as the number of the photometric and kinematic points, the values of the relevant reduced chi-squared, and the corresponding residuals, are listed in Table~\ref{table4}. To evaluate the goodness of the fits, in Table~\ref{table_stat} we compare the values of the reduced photometric and kinematic chi-squared, denoted by $\widetilde{\chi}_{\mathrm{p}}^2$ and $\widetilde{\chi}_{\mathrm{k}}^2$ respectively, with the two-sided 90$\%$ confidence interval (CI), calculated with respect to the reduced $\chi^2$-distribution, characterized by the appropriate number of degrees of freedom (see App.~\ref{App.B.Stat} for details).

\begin{figure*}
\centering
\includegraphics[width=0.95\textheight, angle=270]{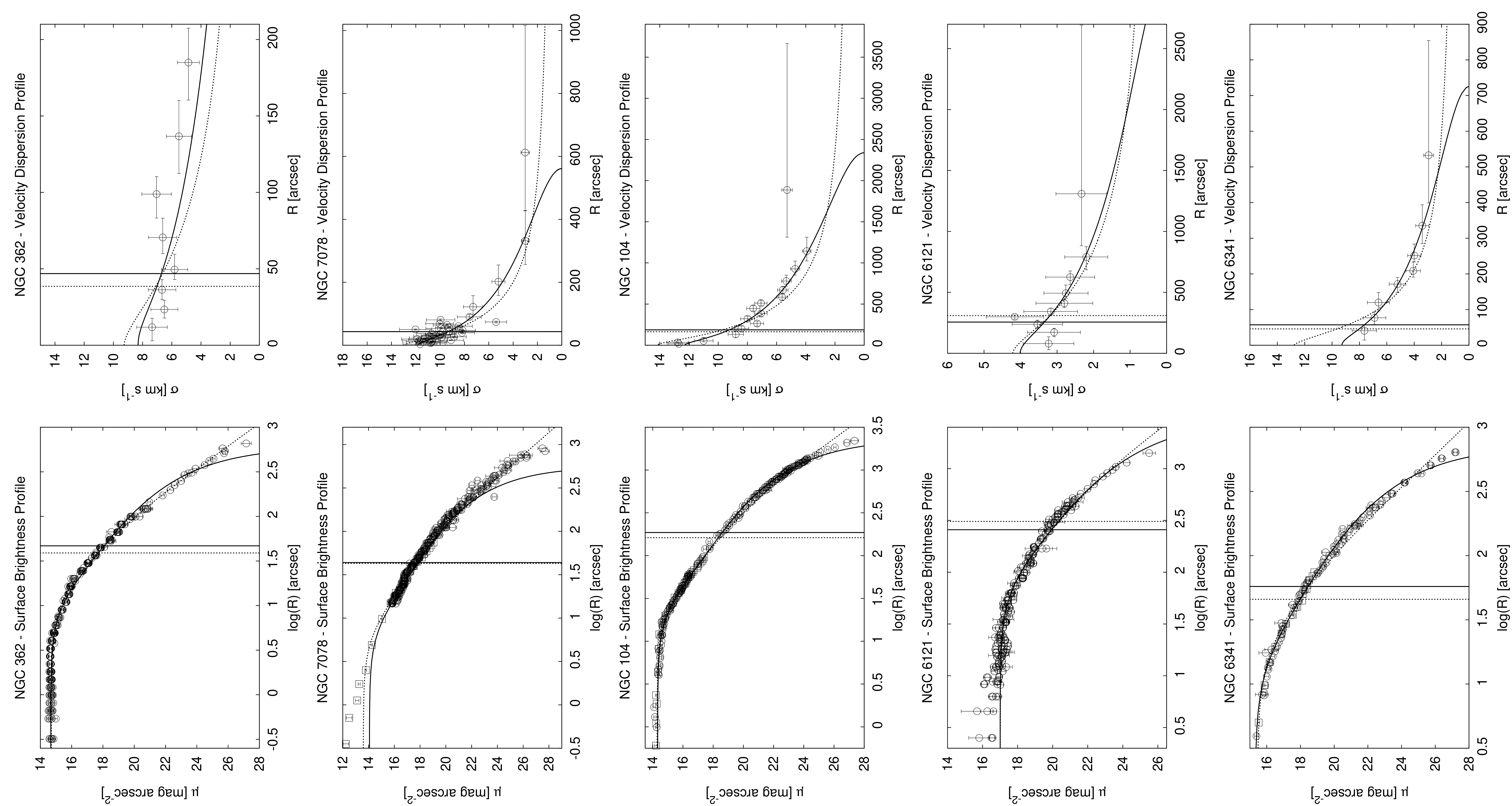}
\caption{Fits by King models and anisotropic $f^{(\nu)}$ models to the surface brightness profiles and to the line-of-sight velocity dispersion profiles of relaxed globular clusters. In all panels, solid lines correspond to the King-model fits, dotted lines to $f^{(\nu)}$-model fits; the vertical solid line marks the position of the King model projected half-light radius, $R_{\mathrm{e}}$, the dotted one the position of the $f^{(\nu)}$ model projected half-light radius, $R_{\mathrm{e}}$. For the surface brightness profiles, the data from \citet{TKD1995} are indicated with circles, the data from other sources (see Sect.~\ref{Subsect.3.1}) with squares. For each data-point, errors are shown as vertical error bars; in the case of the velocity dispersion profile, the horizontal bars indicate the length of the radial bin in which the data-points have been calculated and have no role in the fitting procedure (see Appendix~\ref{App.B}).}
\label{figure1}
\end{figure*}

\begin{figure*}
\centering
\includegraphics[width=\textheight, angle=270]{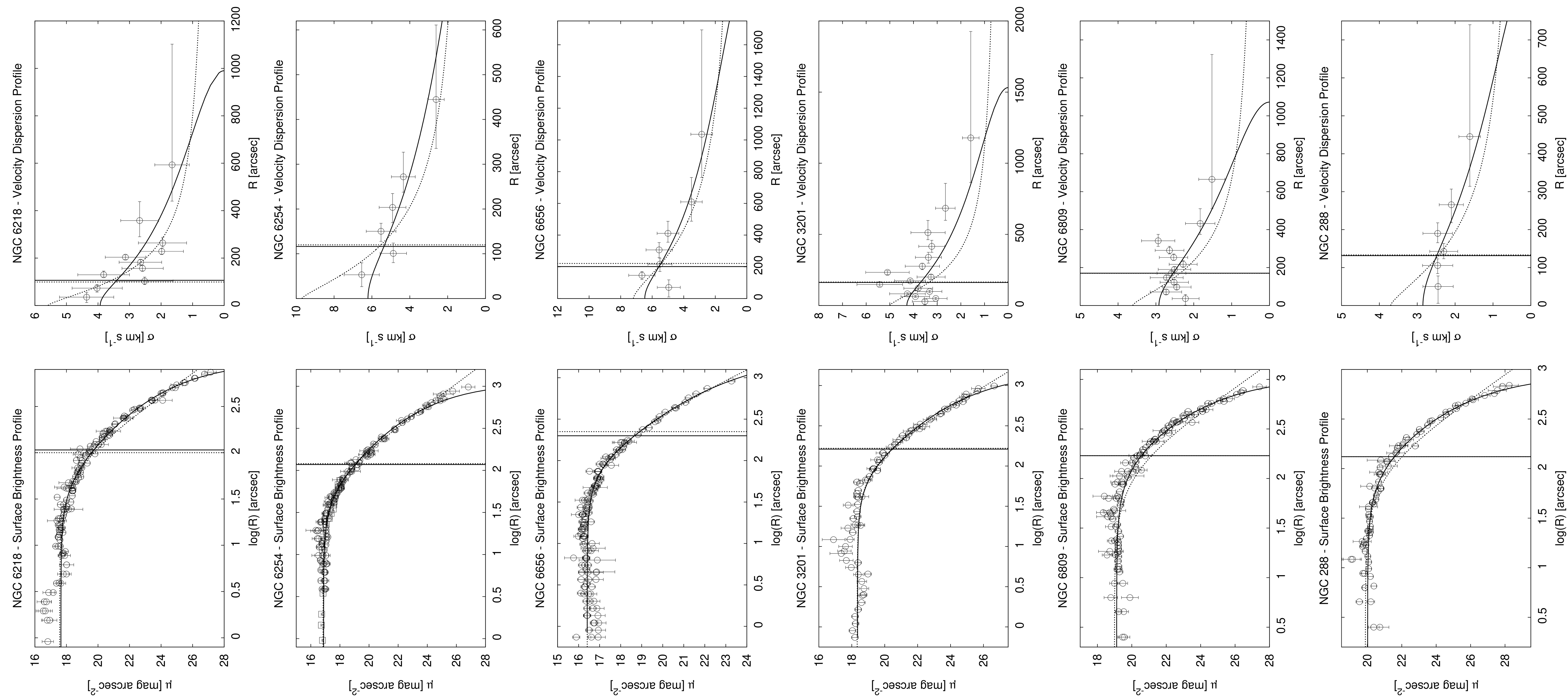}
\caption{Fits by King models and anisotropic $f^{(\nu)}$ models to the surface brightness profiles and to the line-of-sight velocity dispersion profiles of globular clusters in the intermediate relaxation condition, in the same format as in Fig.~\ref{figure1}.}
\label{figure2}
\end{figure*}

\begin{figure*}
\centering
\includegraphics[width=0.375\textheight, angle=270]{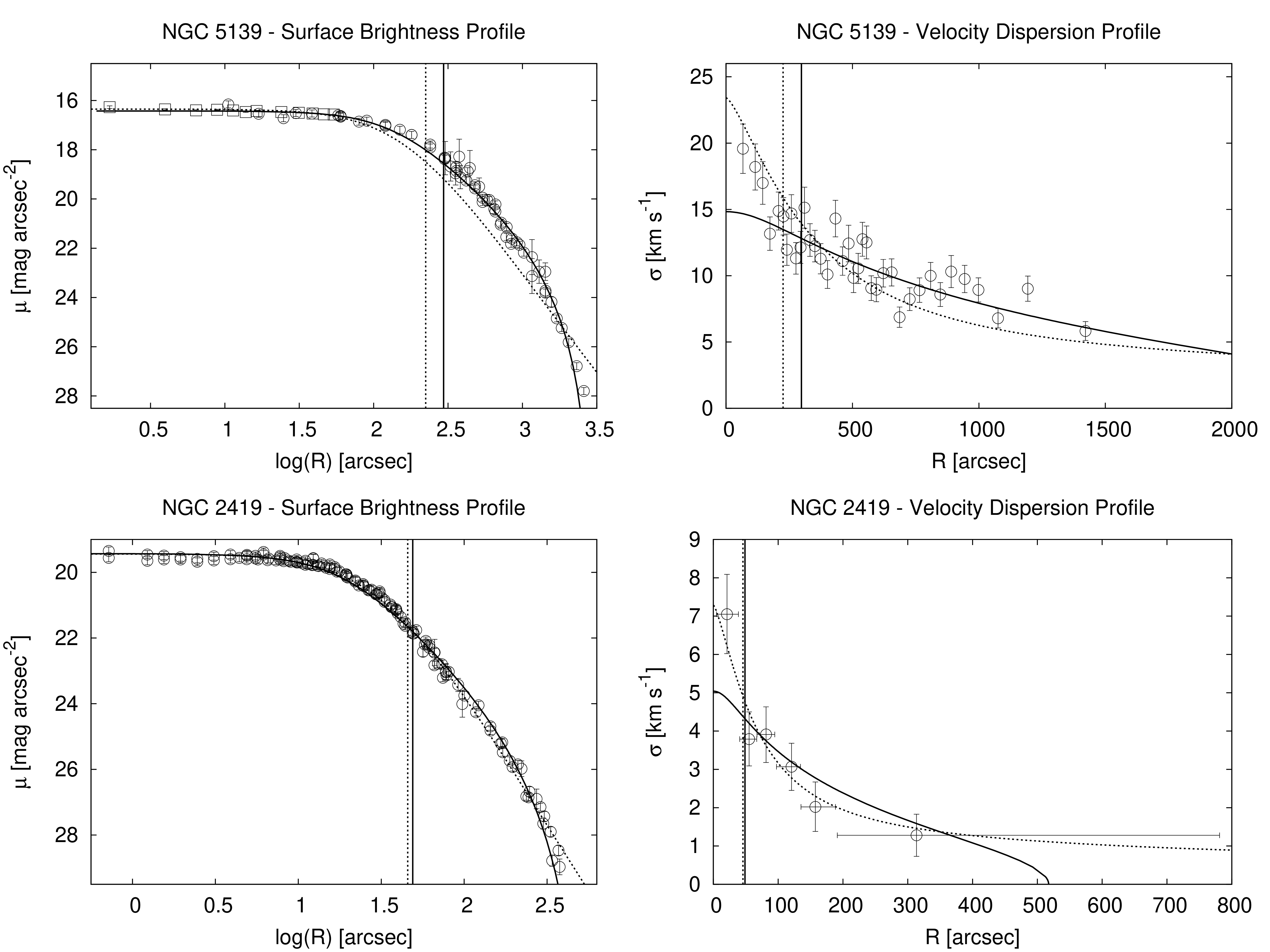}
\caption{Fits by King models and anisotropic $f^{(\nu)}$ models to the surface brightness profiles and to the line-of-sight velocity dispersion profiles of partially relaxed globular clusters, in the same format as in Fig.~\ref{figure1}.}
\label{figure3}
\end{figure*}

The surface brightness and the line-of-sight velocity dispersion profiles determined by the fit procedure for the models, together with the observed profiles for the clusters in the first, second, and third relaxation class, are shown in Figs.~\ref{figure1}, \ref{figure2}, and \ref{figure3}, respectively. In the panels, solid lines correspond to the best-fit King models and dotted lines to the best-fit $f^{(\nu)}$ models. The vertical solid line marks the position of the King model projected half-light radius, the dotted one the position of the $f^{(\nu)}$ model projected half-light radius. For the surface brightness profiles, the data from \citet{TKD1995} are indicated with circles, the data from other sources with squares. For the velocity dispersion data, the horizontal bars indicate the length of the radial bin in which the data-points have been calculated; they do not have a role in determining the fit. For each data-point the errors are shown as vertical error bars. Note that, even if we insisted on selecting clusters with a reasonable number of data inside $R_{\mathrm{e}}$, for about half of the clusters, the kinematic profiles are undersampled in their central region.

\begin{table*}
\caption[]{Quality of the fits.}
\label{table4}
\centering
\begin{tabular}{rrr c cccccc c cccccc}
\hline\hline
 & & & & & \multicolumn{4}{c}{King Models} & & & & \multicolumn{4}{c}{$f^{(\nu)}$ Models} & \\
\cline{5-10}  \cline{12-17}
NGC & $N_{\mathrm{p}}$ & $N_{\mathrm{k}}$ & & $\widetilde{\chi}^2_{\mathrm{p}}$ & $\langle\Delta \mu\rangle$ & $(\Delta \mu)_{\mathrm{max}}$ & $\widetilde{\chi}^2_{\mathrm{k}}$ & $\langle\Delta \sigma\rangle$ & $(\Delta \sigma)_{\mathrm{max}}$ & & $\widetilde{\chi}^2_{\mathrm{p}}$ & $\langle\Delta \mu\rangle$ & $(\Delta \mu)_{\mathrm{max}}$ & $\widetilde{\chi}^2_{\mathrm{k}}$ & $\langle\Delta \sigma\rangle$ & $(\Delta \sigma)_{\mathrm{max}}$\\
(1)  & (2) & (3)& & (4)   & (5)  & (6)  & (7)   & (8)  & (9)  & & (10) & (11) & (12) & (13) & (14) & (15) \\
\hline
104  & 231 & 16 & & 3.487 & 0.41 & 4.68 & 7.411 & 1.03 & 3.39 & &  6.433 & 0.24 & 1.53 & 10.367 & 1.34 & 2.98 \\
288  &  85 &  6 & & 1.251 & 0.30 & 0.98 & 0.442 & 0.21 & 0.33 & &  3.891 & 0.46 & 1.10 &  2.040 & 0.46 & 0.78 \\
362  & 239 &  8 & & 3.113 & 0.58 & 7.09 & 1.307 & 0.99 & 1.74 & &  1.563 & 0.15 & 0.92 &  3.345 & 1.55 & 2.45 \\
2419 & 137 &  6 & & 1.983 & 0.21 & 1.10 & 1.344 & 0.98 & 2.21 & &  1.492 & 0.16 & 0.83 &  0.471 & 0.50 & 0.83 \\
3201 &  80 & 16 & & 1.308 & 0.38 & 1.49 & 1.783 & 0.83 & 1.73 & &  1.289 & 0.36 & 1.48 &  4.005 & 1.21 & 2.29 \\
5139 &  72 & 37 & & 3.750 & 0.36 & 2.08 & 1.974 & 1.73 & 4.90 & & 21.742 & 0.80 & 1.62 &  3.406 & 2.07 & 4.38 \\
6121 & 228 & 10 & & 1.460 & 0.27 & 1.33 & 0.450 & 0.47 & 0.94 & &  1.710 & 0.29 & 1.33 &  0.581 & 0.52 & 0.93 \\
6218 & 143 & 11 & & 1.185 & 0.32 & 1.12 & 0.584 & 0.54 & 0.91 & &  2.663 & 0.40 & 1.12 &  0.765 & 0.59 & 1.10 \\
6254 & 162 &  6 & & 5.046 & 0.37 & 2.75 & 0.606 & 0.48 & 0.65 & &  4.372 & 0.22 & 1.12 &  1.844 & 0.89 & 1.15 \\
6341 & 118 &  8 & & 8.439 & 0.41 & 2.51 & 1.418 & 0.51 & 1.02 & & 20.589 & 0.33 & 1.06 &  2.354 & 1.01 & 2.37 \\
6656 & 143 &  7 & & 1.019 & 0.23 & 0.66 & 0.942 & 0.67 & 1.36 & &  1.056 & 0.23 & 0.66 &  1.699 & 0.89 & 1.83 \\
6809 & 114 & 13 & & 1.165 & 0.32 & 1.04 & 1.103 & 0.40 & 0.96 & &  4.404 & 0.59 & 1.34 &  2.967 & 0.64 & 1.35 \\
7078 & 310 & 35 & & 6.136 & 0.75 & 5.00 & 3.229 & 1.33 & 3.06 & &  3.813 & 0.36 & 1.41 &  1.981 & 1.37 & 3.25 \\
\hline
\end{tabular}
\tablefoot{For each cluster, named in column (1), we provide the number of points in the surface brightness (2) and in the velocity dispersion (3) profile. For King and $f^{(\nu)}$ models, we list: the reduced best-fit photometric chi-squared $\widetilde{\chi}^2_{\mathrm{p}}$ (Col.~(4) and (10)), the mean (Col.~(5) and (11)) and maximum (Col.~(6) and (12)) photometric residuals, the reduced best-fit kinematic chi-squared $\widetilde{\chi}^2_{\mathrm{k}}$ (Col.~(7) and (13)), the mean (Col.~(8) and (14)) and maximum (Col.~(9) and (15)) kinematic residuals.}
\end{table*}

\begin{table*}[ht!]
\caption[]{Two-sided confidence intervals for the reduced $\chi^2$-distribution with $n$ degrees of freedom.}
\label{table_stat}
\centering
\begin{tabular}{r c ccccc c ccccc}
\hline\hline
 & & \multicolumn{5}{c}{Photometric Fits} & & \multicolumn{5}{c}{Kinematic Fits} \\
\cline{3-7}  \cline{9-13}
NGC  & & $n$ & King $\widetilde{\chi}^2_{\mathrm{p}}$ & $f^{(\nu)}$ $\widetilde{\chi}^2_{\mathrm{p}}$ & $\widetilde{\chi}^2_{\mathrm{inf}}$ & $\widetilde{\chi}^2_{\mathrm{sup}}$ & & $n$ & King $\widetilde{\chi}_{\mathrm{k}}^2$ & $f^{(\nu)}$ $\widetilde{\chi}_{\mathrm{k}}^2$ & $\widetilde{\chi}^2_{\mathrm{k,inf}}$ & $\widetilde{\chi}^2_{\mathrm{k,sup}}$ \\
(1)  & & (2)  & (3)    & (4)    & (5)    & (6)   & & (7)  & (8)      & (9)  & (10)  & (11) \\
\hline 
104  & & 229 & 3.49 &  6.43 & 0.85 & 1.16 & & 15 & 7.41 & 10.37 & 0.48 & 1.67 \\
288  & &  83 & 1.25 &  3.89 & 0.76 & 1.27 & &  5 & 0.44 &  2.04 & 0.23 & 2.21 \\
362  & & 237 & 3.11 &  1.56 & 0.85 & 1.16 & &  7 & 1.31 &  3.35 & 0.31 & 2.01 \\
2419 & & 135 & 1.98 &  1.49 & 0.81 & 1.21 & &  5 & 1.34 &  0.47 & 0.23 & 2.21 \\
3201 & &  78 & 1.31 &  1.29 & 0.75 & 1.28 & & 15 & 1.78 &  4.01 & 0.48 & 1.67 \\
5139 & &  70 & 3.75 & 21.74 & 0.74 & 1.29 & & 36 & 1.97 &  3.41 & 0.65 & 1.42 \\
6121 & & 226 & 1.46 &  1.71 & 0.85 & 1.16 & &  9 & 0.45 &  0.58 & 0.37 & 1.88 \\
6218 & & 141 & 1.19 &  2.66 & 0.81 & 1.20 & & 10 & 0.58 &  0.77 & 0.39 & 1.83 \\
6254 & & 160 & 5.05 &  4.37 & 0.82 & 1.19 & &  5 & 0.61 &  1.84 & 0.23 & 2.21 \\
6341 & & 116 & 8.44 & 20.59 & 0.79 & 1.23 & &  7 & 1.42 &  2.35 & 0.31 & 2.01 \\
6656 & & 141 & 1.02 &  1.06 & 0.81 & 1.20 & &  6 & 0.94 &  1.70 & 0.27 & 2.10 \\
6809 & & 112 & 1.17 &  4.40 & 0.79 & 1.23 & & 12 & 1.10 &  2.97 & 0.44 & 1.75 \\
7078 & & 308 & 6.14 &  3.81 & 0.87 & 1.14 & & 34 & 3.23 &  1.98 & 0.64 & 1.43 \\
\hline
\end{tabular}
\tablefoot{For each cluster, named in column (1), separately for the photometric and the kinematic fits, we provide the number of degrees of freedom of each fit (Col.~(2) and (7)), the reduced best-fit chi-squared for King (Col.~(3) and (8)) and for $f^{(\nu)}$ (Col.~(4) and (9)) models, and the lower (Col.~(5) and (10)) and upper (Col.~(6) and (11)) boundaries of the two-sided 90$\%$ confidence level interval for the reduced $\chi^2$-distribution with $n$ degrees of freedom.}
\end{table*}

In the following part of this subsection we will try to give a general assessment of the quality of the fits in the various cases. As a general rule, for a given cluster the family of models yielding the lowest values of the best-fit photometric and kinematic chi-squared is preferred. In practice, the objective properties of the fits are best obtained by checking directly the values provided in Tables~\ref{table4} and \ref{table_stat} and by inspection of Figs.~\ref{figure1}-\ref{figure3}. Table~\ref{table4} shows that generally, for a given cluster, $\widetilde{\chi}^2_{\mathrm{p}} > \widetilde{\chi}^2_{\mathrm{k}}$, because the photometric profiles are characterized by a larger number of data-points with reported smaller error-bars. Table~\ref{table_stat} shows that NGC 6656 is the only cluster for which the values of $\widetilde{\chi}^2_{\mathrm{p}}$ and $\widetilde{\chi}^2_{\mathrm{k}}$ for the two models are inside the relevant 90$\%$ CI. Within the King modeling, four clusters have $\widetilde{\chi}^2_{\mathrm{p}}$ inside the 90$\%$ CI and nine have $\widetilde{\chi}^2_{\mathrm{k}}$ inside the 90$\%$ CI. A visual inspection of the fits (in particular, see the kinematic fits for NGC 362, NGC 3201, NGC 6121, NGC 6656, NGC 6809 and the inner photometric profiles of NGC 3201, NGC 6121, NGC 6218, NGC 6656, and NGC 7078) also suggests that some systematic trends are clearly missed by the two families of models used.

Within the class of relaxed globular clusters, for NGC 362 and NGC 7078 it is evident that the King models cannot reproduce the observed surface brightness profiles, and that the $f^{(\nu)}$ models perform better, \textit{especially for describing the outer parts of the cluster}. As to the observed velocity dispersion profiles, we see that for NGC 7078 the $f^{(\nu)}$ profile is formally more adequate (at the 99.9$\%$ confidence level, which, in the following, we denote by CL), while for NGC 362 the King profile is the closer to the observations (at the 90$\%$ CL). We should recall that these two globular clusters are flagged in the Harris catalogue as post-core-collapse clusters, although the value of the concentration parameter C is smaller than 2.5. Indeed, the observations indicate that there are some processes that cannot be captured by King models (with particular reference to the shallow cusp in the photometric profile of NGC 7078). By looking at the plots in Fig.~\ref{figure1}, NGC 104 and NGC 6341 appear to have both observed surface brightness and velocity dispersion profiles well represented by King models (for NGC 6341 the King $\widetilde{\chi}^2_{\mathrm{k}}$ falls within the 90$\%$ CI), in spite of the fact that, from Table~\ref{table_stat}, the fits cannot be considered entirely satisfactory. Curiously, NGC 6121 is equally well described by the two models; the values of the reduced chi-squared are slightly lower for the King model, but in practice the quality of the two fits is similar. We notice that for NGC 104 and NGC 6121 it is particularly evident that the last point in the velocity dispersion profile is significantly higher than expected by both models. A partial explanation of this fact could be the very large extent of the radial interval in which it is calculated. Therefore, it is important to obtain more velocity data-points in the outer regions to clarify this issue.


\begin{table*}[ht!]
\caption[]{Derived structural properties.}
\label{table5}
\centering
\begin{tabular}{r c ccccccc c ccccccc}
\hline\hline
 & & & \multicolumn{5}{c}{King Models} & & & & \multicolumn{5}{c}{$f^{(\nu)}$ Models} & \\
\cline{3-9}  \cline{11-17}
NGC  & & $c$ & $R_{\mathrm{c}}$ & $r_{\mathrm{M}}$ & $r_{\mathrm{tr}}$  & $M$ & $M/L$   & $\rho_{\mathrm{0}}$ & & $R_{\mathrm{c}}$ & $r_{\alpha}/r_{\mathrm{M}}$ & $r_{\mathrm{M}}$ & $r_{\alpha}$ & $M$ & $M/L$ & $\rho_{\mathrm{0}}$\\
(1)  & & (2)  & (3)    & (4)    & (5)    & (6)    & (7)  & (8)      & & (9)   & (10)             & (11)  & (12)         & (13)& (14)  & (15)    \\
\hline 
104  & & 2.00 &  22.60 &  2.57 & 2335.74 &  7.181 & 1.34 & 5.011 & & 23.25 & 1.783 &  4.82 &  8.59 &  8.047 & 1.50 & 5.094  \\
288  & & 0.99 &  79.49 &  7.51 &  896.85 &  0.740 & 1.88 & 2.043 & & 58.24 & 0.866 &  7.53 &  6.53 &  0.850 & 2.18 & 2.427 \\
362  & & 1.77 &   9.88 &  2.65 &  605.79 &  1.867 & 1.05 & 4.826 & & 11.34 & 1.786 &  2.13 &  3.80 &  1.828 & 1.03 & 4.761 \\
2419 & & 1.42 &  18.58 & 19.56 &  516.97 &  7.843 & 1.72 & 1.874 & & 20.01 & 0.958 & 23.97 & 22.95 & 10.912 & 2.40 & 2.081 \\
3201 & & 1.30 &  71.58 &  3.85 & 1532.62 &  1.131 & 1.91 & 3.012 & & 72.31 & 0.915 &  5.14 &  4.70 &  1.088 & 1.86 & 3.026 \\
5139 & & 1.32 & 127.68 &  7.51 & 2861.08 & 26.446 & 1.93 & 3.537 & & 98.89 & 0.978 &  7.46 &  7.30 & 35.427 & 2.58 & 4.049 \\
6121 & & 1.62 &  71.31 &  2.62 & 3144.17 &  0.654 & 1.10 & 3.656 & & 72.45 & 1.868 &  4.40 &  8.23 &  0.750 & 1.26 & 3.661 \\
6218 & & 1.28 &  47.81 &  2.47 &  982.45 &  0.614 & 1.96 & 3.306 & & 43.45 & 0.890 &  3.03 &  2.69 &  0.786 & 2.52 & 3.577 \\
6254 & & 1.32 &  50.02 &  2.48 & 1125.66 &  1.532 & 1.61 & 3.741 & & 47.97 & 0.564 &  3.35 &  1.89 &  2.161 & 2.67 & 4.003  \\
6341 & & 1.69 &  14.18 &  2.12 &  724.28 &  2.866 & 1.83 & 4.638 & & 17.12 & 1.523 &  2.43 &  3.70 &  3.956 & 2.53 & 4.691  \\
6656 & & 1.38 &  80.92 &  3.13 & 2057.81 &  2.081 & 1.11 & 3.636 & & 82.71 & 1.523 &  4.53 &  6.89 &  2.337 & 1.24 & 3.652 \\
6809 & & 0.92 & 110.09 &  5.90 & 1072.17 &  0.604 & 1.12 & 2.214 & & 74.28 & 0.867 &  5.83 &  5.06 &  0.627 & 1.16 & 2.628 \\
7078 & & 1.86 &   7.51 &  1.70 &  560.55 &  3.976 & 1.12 & 5.207 & &  6.27 & 1.793 &  2.92 &  5.24 &  4.056 & 1.14 & 5.420  \\
\hline
\end{tabular}
\tablefoot{For each cluster, listed in column (1), for the King models, we provide (2) the concentration index $c = \log(r_{\mathrm{tr}}/r_0)$ (see Eqs. (\ref{r0}), (\ref{c})) and (5) the truncation radius $r_{\mathrm{tr}}$, in arcsec; for the $f^{(\nu)}$ models, in Col.~(10) the ratio between the anisotropy and the half-mass radius and in Col.~(12) the anisotropy radius $r_{\alpha}$, defined as $\alpha(r_{\alpha})=1$ (see Eq. (\ref{alpha}) for the definition of $\alpha$), in pc. For both King and $f^{(\nu)}$ models, we list: the core radius $R_{\mathrm{c}}$ (defined in the standard way) in arcsec (Col.~(3) and (9)), the intrinsic half-mass radius $r_{\mathrm{M}}$, in pc (Col.~(4) and (11), respectively), the total mass $M$ of the cluster (Col.~(6) and (13)) expressed in units of $10^5$ M$_{\sun}$, the V band mass-to-light ratio in solar units (Col.~(7) and (14)) and the logarithm of the central mass density $\rho_{\mathrm{0}}$ in M$_{\sun}$~pc$^{-3}$ (Col.~(8) and (15)).}
\end{table*}

Best-fit profiles for globular clusters in the intermediate relaxation class are shown in Fig.~\ref{figure2}. For NGC 288, NGC 6218, and NGC 6809 King models provide a better fit to both the surface brightness and the velocity dispersion profile (at the 90$\%$ CL in all cases). For the other globular clusters in this relaxation class the results are less sharp. In fact, for NGC 3201 and NGC 6254 the surface brightness profiles are formally better reproduced by $f^{(\nu)}$ models (for the first, at 95$\%$ CL), while the corresponding velocity dispersion profiles are formally better described by King models (at the 95$\%$ and 90$\%$ CL, respectively). For NGC 6656 both the surface brightness and the velocity dispersion profile are approximately equally well reproduced by the two families of models (at the 90$\%$ CL in all cases). For NGC 3201, NGC 6656, and NGC 6809 the velocity dispersion profiles have an irregular shape in the central regions: even if King models formally perform better than $f^{(\nu)}$ models, they are unable to reproduce the observations. We tried to choose a different binning for the data and we found that this irregularity does not depend on the way in which the observed velocity dispersion profile is constructed from the available data-set. Clearly, more data are necessary in order to obtain a more convincing description of these systems.

\begin{figure*}[ht!]
\centering
\includegraphics[height=17cm, angle=270]{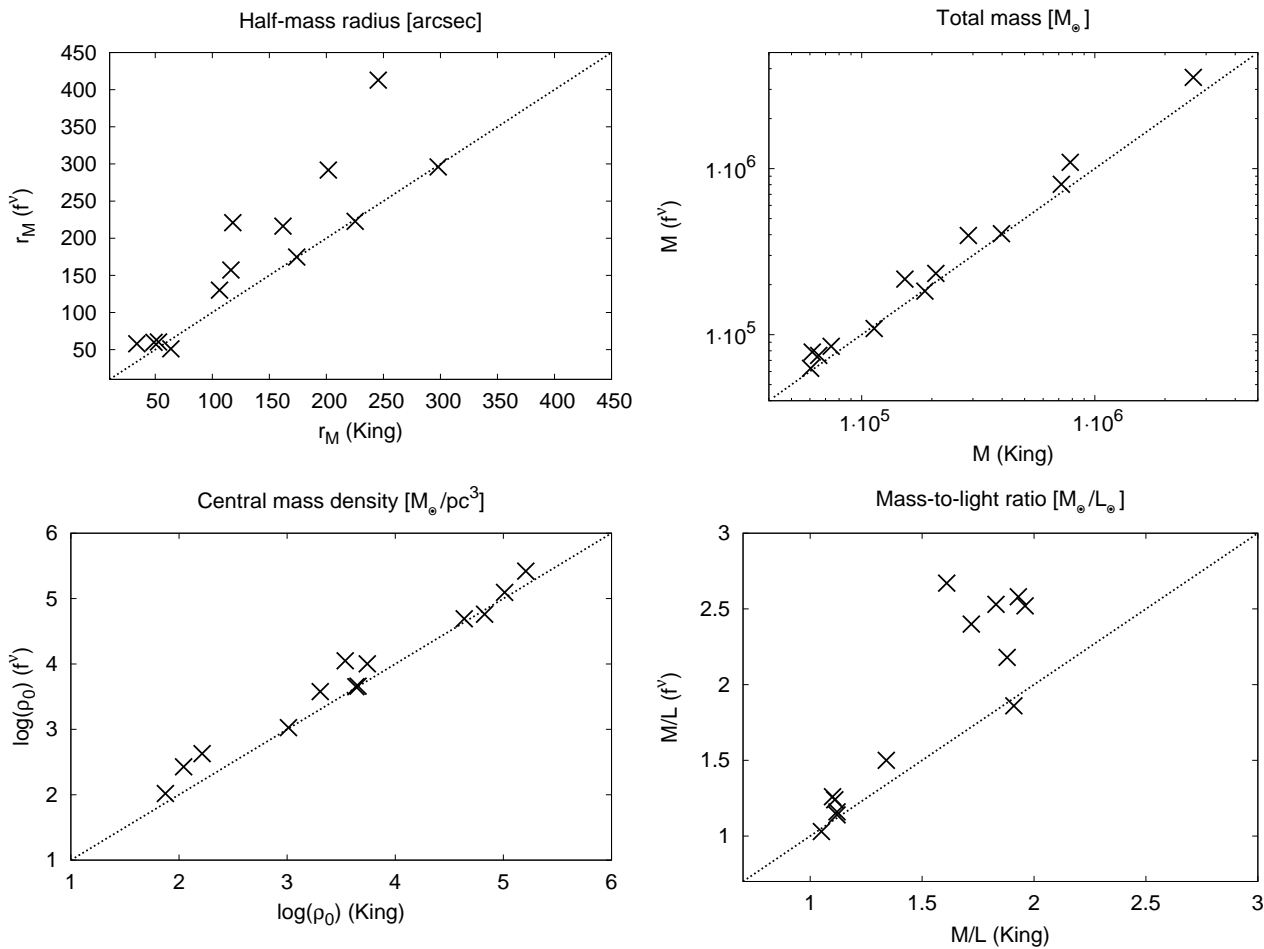}
\caption{Comparison between the values of the structural parameters derived from King and $f^{(\nu)}$ best-fit models. The top left panel shows, as crosses, the values of half-mass radii, the top right panel the total masses, the bottom left panel the central mass density and the bottom right panel the mass-to-light ratios. In each panel, the x-axis refers to the values calculated with King models, and the y-axis to those calculated with $f^{(\nu)}$ models; the dotted diagonal line denotes the identity relation. The intrinsic quantity $r_{\mathrm{M}}$ would be best given in parsecs, but it is reported here in arcseconds for easier comparison with the observed profiles and with the projected half-light radius $R_{\mathrm{e}}$ represented in the previous figures.}
\label{figure4}
\end{figure*}

In Fig.~\ref{figure3}, we show the best-fit profiles for partially relaxed globular clusters. For these two clusters we see discordant results: for NGC 2419 $f^{(\nu)}$ models are more adequate for describing the data (at the 99.99$\%$ CL and 90$\%$ CL for the photometric and kinematic fit, respectively), while for NGC 5139 King models provide a better fit to the observed profiles. Even if formally King models perform better in describing the kinematic profiles of NGC 5139, they do not provide a satisfactory description of the kinematics of the central parts of the cluster (see Fig.~\ref{figure3}); in this respect, the $f^{(\nu)}$ models give a better representation of the inner kinematics.

To summarize, we found that, as expected, $f^{(\nu)}$ models tend to perform globally better than King models for the least relaxed globular cluster of our sample, NGC 2419. For NGC 2419, the good performance of $f^{(\nu)}$ models might correspond to the partial relaxation condition of the cluster, consistent with the physical picture that motivates the definition of the $f^{(\nu)}$ models, as outlined in the Introduction. In addition, for three clusters in the second relaxation class (NGC 3201, NGC 6254,and NGC 6656), $f^{(\nu)}$ models are competitive with King models. Furthermore, $f^{(\nu)}$ models can describe well a relatively steep central slope of the velocity dispersion profile, even when the corresponding photometric profile is cored, while King models and other isotropic truncated models (such as Wilson models) are unable to reproduce this kinematical behavior. This fact is evident from the kinematic fits for NGC 2419, NGC 5139 and for possibly one cluster in the intermediate relaxation condition (NGC 6218).

As far as the behavior of the photometric profiles at large radii is concerned, we see that, especially for NGC 104, NGC 362, NGC 6254, NGC 6341, and NGC 7078 King models do not provide a good description of the truncation, as noted in a number of previous studies (in particular, see \citealp{MLvdM2005}; \citealp{JordiGrebel2010}; \citealp{Kuepper2010}, in which the outermost parts of the surface brightness profiles are appropriately modeled by using N-body simulations). In some cases, the observed profile falls between the King and the $f^{(\nu)}$ profiles. This suggests that truncated $f^{(\nu)}$ models might behave systematically better than King models for describing these stellar systems. To a large extent, the modification by truncation in phase space of the anisotropic (nontruncated) $f^{(\nu)}$ models is complementary to the generalization of the isotropic (truncated) King models to models characterized by anisotropic pressure, that is, the so-called Michie-King models (in which the truncated Maxwellian is associated with the anisotropic factor of the Eddington models; see \citealp{Michie1963}, \citealp{GunnGriffin1979}). At this stage, it would be interesting to compare the behavior of the outer photometric and kinematic profiles of the two families of anisotropic truncated models, to evaluate the interplay between truncation and pressure anisotropy in the two different cases. Of course, the simple physical picture offered by King, $f^{(\nu)}$, and Michie-King models still suffers from a number of limitations, as discussed in the Introduction. Such a simple picture is bound to fail in the modeling of clusters in which core collapse has taken place. Therefore, it is not surprising that clusters such as NGC 362 and NGC 7078, even if they belong to the first relaxation class, are not described by
isotropic King models as well as expected.

\begin{figure*}
\centering
\includegraphics[height=17cm, angle=270]{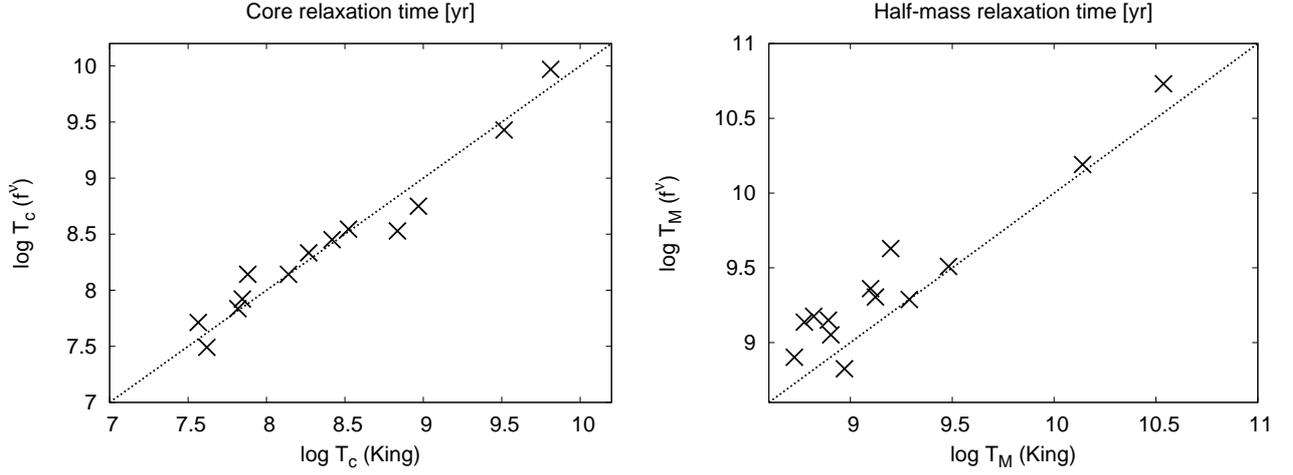}
\caption{Comparison between the values of the relaxation times derived from King and $f^{(\nu)}$ best-fit models. The left panel shows the values of core relaxation times, the right panel the half-mass relaxation times. The format is the same as in Fig.~\ref{figure4}.}
\label{figure5}
\end{figure*}

In closing, we wish to reiterate that, in general, the kinematic fits are crucial to assess if a model is actually able to describe a given globular cluster. Unfortunately, the observed velocity dispersion profiles are generally less accurate and less reliable, with respect to the surface brightness profiles; not only the outer parts (radii close to the truncation radius), but also the inner parts (inside the half-light radius) are often not well sampled as would be desired. The present study confirms that in the future it would be desirable to acquire new and better kinematic data.

\subsection{King models vs. $f^{(\nu)}$ models}
\label{Subsect.4.2}

The values of the relevant structural parameters derived from the best-fit models are presented in Table~\ref{table5}. To compare quantitatively the properties of the best-fit models selected in the two families of models for each globular cluster, we can correlate the values of the derived parameters, such as the half-mass radius $r_{\mathrm{M}}$, the total mass $M$, the central mass density $\rho_0$ and the mass-to-light ratio $M/L$. These correlations are illustrated in Fig.~\ref{figure4}.

For the majority of the globular clusters considered in our sample, the values of the half-mass radius from $f^{(\nu)}$ models are larger than those obtained from King models; only for NGC 362 the opposite is true. For NGC 288, NGC 5139 and NGC 6809 the values calculated from the two models basically coincide. By comparing the values of the total mass and of the central mass density calculated with the two families of models, we see that for the whole sample (with few exceptions) the values calculated with $f^{(\nu)}$ models are larger than those calculated with King models; this fact is not at all surprising, since the $f^{(\nu)}$ models are not truncated. As to the mass-to-light ratios, we see that there is not a tight correlation between the values calculated with King and $f^{(\nu)}$ models, the latter being almost always larger. Similar trends are noted also in the structural properties derived from (isotropic) models characterized by a more spatially extended truncation, such as the Wilson models (see \citealp{MLvdM2005}).

In Table~\ref{trelax}, we list the values of the core and half-mass relaxation times calculated by using the two best-fit dynamical models for each globular cluster; the clusters are listed in order of increasing King core relaxation times, and the separation in three classes of relaxation here adopted is marked with horizontal lines. Figure~\ref{figure5} illustrates the correlation between the values of these parameters.

\begin{table}
\caption[]{Core and half-mass relaxation times for the best-fit King and $f^{(\nu)}$ models.}
\label{trelax}
\centering
\begin{tabular}{r c cc c cc}
\hline\hline
& & \multicolumn{2}{c}{King Models} & & \multicolumn{2}{c}{$f^{(\nu)}$ Models} \\
NGC  & & $\log{T_{\mathrm{c}}}$ & $\log{T_{\mathrm{M}}}$ & & $\log{T_{\mathrm{c}}}$ & $\log{T_{\mathrm{M}}}$ \\
(1)  & & (2)         & (3)         & & (4)         & (5)         \\
\hline
362  & & 7.565 &  8.971 & & 7.713 &  8.825 \\
7078 & & 7.620 &  8.820 & & 7.491 &  9.176 \\
6121 & & 7.818 &  8.774 & & 7.836 &  9.137 \\
104  & & 7.846 &  9.198 & & 7.922 &  9.629 \\
6341 & & 7.881 &  8.904 & & 8.142 &  9.052 \\
\hline
6218 & & 8.140 &  8.725 & & 8.142 &  8.902 \\
6254 & & 8.269 &  8.892 & & 8.334 &  9.150 \\
6656 & & 8.418 &  9.099 & & 8.451 &  9.361 \\
3201 & & 8.523 &  9.123 & & 8.545 &  9.305 \\
6809 & & 8.835 &  9.289 & & 8.529 &  9.288 \\
288  & & 8.969 &  9.482 & & 8.750 &  9.509 \\
\hline
5139 & & 9.515 & 10.140 & & 9.429 & 10.191 \\
2419 & & 9.812 & 10.537 & & 9.970 & 10.731 \\
\hline
\end{tabular}
\end{table}

When considering the core relaxation times (calculated according to Eq.~(10) of \citet{Djorgovski1993}, as in the Harris 2010 catalog), we see that the original division in the relaxation classes proposed in this paper on the basis of the values listed in the Harris catalog is confirmed. The only exception is NGC 6341: according to the value of the core relaxation time calculated with its best-fit $f^{(\nu)}$ model, this cluster should belong to the second class, rather than to the first. When we list globular clusters according to increasing core relaxation times estimated with the models identified in this paper, we see that the order of relaxed globular clusters changes with respect to that of Table~\ref{table1}. However, there is a general agreement between values of these quantities calculated with the two families of models.

\begin{table*}
\caption[]{Comparison between the values of structural parameters from best-fit King models found in this paper and in previous studies. See Table~\ref{table3} for the description of column entries.}
\label{table6}
\centering
\begin{tabular}{ccccccccccccc}
\hline\hline
NGC & Ref. &  $\Psi$ &  $r_{\mathrm{0}}$  & $\mu_{\mathrm{0}}$ &   $V$   &   & NGC & Ref. &  $\Psi$ &  $r_{\mathrm{0}}$  & $\mu_{\mathrm{0}}$ &   $V$ \\
\cline{1-6} \cline{8-13}
104  & (0)  &  8.58 &  23.09 & 14.33 & 12.27 &   &  288 & (0)  &  4.82 &  91.03 & 20.02 &  2.85 \\
     & (1)  &  9.33 &  15.25 & 14.42 & 11.5  &   &      & (1)  &  4.33 &  94.05 & 20.00 &  2.9  \\
     & (2)  &  8.70 &  22.84 & 14.42 & $\ldots$ & &     & (2)  &  4.65 &  98.41 & 20.00 & $\ldots$ \\
     & (3)  &  8.6  &  23.19 & 14.36 & 15.27 &   &      & (3)  &  4.8  &  92.87 & 20.00 &  2.79 \\
     & (4)  &  8.81 &  22.01 & 14.38 & 11.0  &   &      & (4)  &  4.80 &  92.79 & 20.05 &  2.9  \\
362  & (0)  &  7.80 &  10.21 & 14.66 &  8.31 &   & 2419 & (0)  &  6.62 &  19.70 & 19.43 &  5.04 \\
     & (1)  &  8.93 &   7.19 & 14.79 &  6.4  &   &      & (1)  &  6.55 &  19.88 & 19.77 &  3.0  \\
     & (2)  &  6.80 &  10.79 & 14.79 & $\ldots$ & &     & (2)  &  6.55 &  22.19 & 19.77 & $\ldots$ \\
     & (3)  &  7.9  &  10.05 & 14.66 & 11.12 &   &      & (3)  &  6.5  &  20.60 & 19.44 &  5.32 \\
     & (4)  &  7.76 &  11.16 & 14.80 &  6.4  &   &      & (4)  &  6.44 &  20.47 & 19.67 &  4.0  \\
3201 & (0)  &  6.17 &  76.99 & 18.35 &  4.28 &   & 5139 & (0)  &  6.27 & 136.94 & 16.42 & 14.83 \\
     & (1)  &  6.17 &  86.22 & 18.96 &  5.2  &   &      & (1)  &  5.77 & 157.85 & 16.81 & 16.0  \\
     & (2)  &  6.21 &  93.55 & 18.96 & $\ldots$ & &     & (2)  &  5.94 & 168.01 & 16.81 & $\ldots$ \\
     & (3)  &  6.1  &  77.68 & 18.30 &  4.21 &   &      & (3)  &  6.2  & 141.20 & 16.44 & 13.75 \\
     & (4)  &  6.14 &  83.98 & 19.00 &  5.0  &   &      & (4)  &  6.21 & 152.73 & 16.81 & 16.8  \\
6121 & (0)  &  7.32 &  74.56 & 17.00 &  4.01 &   & 6218 & (0)  &  6.11 &  51.56 & 17.65 &  3.93 \\
     & (1)  &  7.24 &  45.38 & 17.88 &  4.2  &   &      & (1)  &  6.55 &  37.57 & 18.00 &  4.5  \\
     & (2)  &  7.20 &  52.37 & 17.88 & $\ldots$ & &     & (2)  &  6.48 &  42.38 & 18.00 & $\ldots$ \\
     & (3)  &  7.4  &  72.57 & 16.94 &  5.25 &   &      & (3)  &  6.1  &  51.74 & 17.62 &  4.66 \\
     & (4)  &  7.40 &  72.42 & 17.95 &  4.0  &   &      & (4)  &  6.33 &  50.70 & 18.10 &  4.5  \\
6254 & (0)  &  6.26 &  53.63 & 16.88 &  6.21 &   & 6341 & (0)  &  7.54 &  14.72 & 15.31 &  9.28 \\
     & (1)  &  6.55 &  48.45 & 17.70 &  6.6  &   &      & (1)  &  8.22 &  11.27 & 15.46 &  5.9  \\
     & (2)  &  6.55 &  54.49 & 17.70 & $\ldots$ & &     & (2)  &  7.92 &  14.56 & 15.46 & $\ldots$ \\
     & (3)  &  6.5  &  49.41 & 16.85 &  6.17 &   &      & (3)  &  7.5  &  16.15 & 15.61 &  8.71 \\
     & (4)  &  6.48 &  49.19 & 17.70 &  6.6  &   &      & (4)  &  7.50 &  16.20 & 15.47 &  6.0  \\
6656 & (0)  &  6.47 &  86.18 & 16.41 &  6.47 &   & 6809 & (0)  &  4.44 & 129.11 & 19.12 &  2.92 \\
     & (1)  &  6.17 &  83.19 & 17.40 &  9.0  &   &      & (1)  &  3.17 & 192.94 & 19.40 &  4.9  \\
     & (2)  &  6.21 &  91.41 & 17.40 & $\ldots$ & &     & (2)  &  3.53 & 215.54 & 19.40 & $\ldots$ \\
     & (3)  &  6.5  &  85.15 & 16.38 &  8.56 &   &      & (3)  &  4.5  & 126.41 & 19.07 &  3.73 \\
     & (4)  &  6.48 &  84.96 & 17.42 &  7.8  &   &      & (4)  &  4.49 & 126.22 & 19.36 &  4.0  \\
7078 & (0)  &  8.09 &   7.72 & 14.07 & 11.83 &   &      &      &       &        &       &       \\
     & (1)  & 12.32 &   1.96 & 14.21 & 12.0  &   &      &      &       &        &       &       \\
     & (2)  & 10.75 &   4.09 & 14.21 & $\ldots$ & &     &      &       &        &       &       \\
     & (4)  &  9.72 &   8.49 & 14.21 & 13.5  &   &      &      &       &        &       &       \\
\hline
\end{tabular}
\tablefoot{For NGC 7078 the entry marked with (3) is missing: the authors of this work decided to exclude this  globular cluster from their analysis because it is core-collapsed. In references (2) and (4) only the core radii are given; therefore we used the values of $R_{\mathrm{c}}$ and $\Psi$ to calculate $r_{\mathrm{0}}$ (see App.~\ref{App.A.1}).}
\tablebib{(0) This work; (1) \citealp{PryorMeylan1993}; (2) \citealp{TKD1995}; (3) \citealp{MLvdM2005}; (4) \citealp{HarrisCat2010}.}
\end{table*}

To calculate the half-mass relaxation times we followed the definition of Eq.~(5) in \citet{Spitzer1971}, which is  based on the half-mass radius. We see that the values from $f^{(\nu)}$ models are larger than those from King models, except for NGC 362 and NGC 6809 (for the latter cluster, the estimated values are approximately equal). In contrast with the case of the core relaxation times, the values of the half-mass relaxation time turn out to be more model-dependent. Such dependence is likely to be due to the differences between the density profiles of the two families of models, which, in general, are less significant in the central regions and become more evident at radii larger than the half-mass radius. The introduction of a truncation for $f^{(\nu)}$ models would also lead to different values of these parameters. We notice that usually the half-mass relaxation time is calculated by inserting in the relevant definition directly the (projected) half-light radius (see, for example, \citealp{HarrisCat2010} and \citealp{MLvdM2005}). By following the same procedure, we found that the values of the half-mass relaxation times are less model dependent, and closer to the values listed in the Harris catalog.

\subsection{Comparison with previous studies}
\label{Subsect.4.3}

In the case of King models, it is interesting to compare the values of the structural parameters found in the present paper with those obtained in previous studies, as summarized in Table~\ref{table6}. By combining the formal errors from both analyses, the comparison of this paper with \citet{MLvdM2005} is the most significant because their values result from a fitting procedure which is similar to the one we have followed; for this reason, it is also, to some extent, the most surprising. We recall that in this paper the errors on the parameters can be found in Table~\ref{table3}. \citet{MLvdM2005} list the formal errors on the parameters in their Tables~10 and 12. No similarly detailed comparison with other papers could be made, because in general error analysis is not provided.

We notice that for ten globular clusters (all but NGC 2419, NGC 6254, and NGC 7078) our values of $\Psi$ agree, within the errors, with the values of \citet{MLvdM2005}; for the scale radius, agreement is found for eight objects (all but NGC 2419, NGC 6121, NGC 6254, NGC 6341, and NGC 7078). We may argue that the discrepancy is partly due to the fact that our surface brightness profiles usually contain a larger number of data-points with respect to those of the cited paper (see Sect.~\ref{Subsect.3.1}), even when we do not merge the original profiles by \citet{TKD1995} with those of other sources.

The discrepancies between the values of the central surface brightness derived in this paper and those resulting from previous studies are primarily due the correction for the extinction, usually neglected, introduced in our analysis, following \citet{MLvdM2005}. In fact, such discrepancies are not severe; only three objects (NGC 6341, NGC 6809, and NGC 7078) have values which are not consistent, within the errors, with respect to values determined in previous studies. The reason for this is the fact that we added to these profiles the more recent and more accurate data from \citet{Noyola2006}.

The structural parameter for which the differences between the values obtained in our analysis and those in the literature are the most relevant is the central line-of-sight velocity dispersion. For NGC 288, NGC 2419, NGC 3201, NGC 6121, NGC 6254, and NGC 7078 we see that our values agrees, within the errors, with at least one of the values found in the literature (only four of them with \citealp{MLvdM2005}). This discrepancy is not at all surprising, because the kinematic data are the most uncertain. This fact gives one further argument for the need for more numerous and more accurate kinematic data for these systems.

\subsection{Central slope of the photometric and kinematic profiles}
\label{Subsect.4.4}

When considering the surface brightness profiles of our selected globular clusters, we note that there are cases in which the observed photometric profiles deviate from the calculated ones at small radii. In particular, by focusing on the innermost regions of the profiles, we see that this is the case for NGC 6121, NGC 6218, and NGC 7078, for which the models are underluminous, and NGC 288, NGC 6656, and NGC 6809, in which the models are overluminous, and NGC 3201, for which the observed central surface brightness appears to oscillate; in spite of these local discrepancies, the global values of the statistical indicators may be satisfactory (see Table~\ref{table_stat}).

As far as the velocity dispersion profiles are concerned, in four cases (NGC 288, NGC 3201, NGC 6121, and NGC 6656), both models overpredict the central data-points, while, as mentioned in Sect.~\ref{Subsect.4.1}, four clusters (NGC 2419, NGC 5139, and, possibly, NGC 6218 and NGC 6254) show a relatively large gradient of the profile in the central regions. At variance with the study of elliptical galaxies, the kinematic profiles of globular clusters are often undersampled inside the half-light radius.

Recently, the central cusps in the observed photometric and kinematic profiles of some globular clusters have been interpreted as clues of the presence of an intermediate-mass black hole (IMBH) in the center of the system (see the analytical model by \citealp{BW1976} and the N-body simulations by \citealp{Baumgardt2005} and \citealp{NoyolaBaumgardt2011}), and a variety of dynamical models (either defined from distribution functions or as solutions of the Jeans equations), have been used in order to constrain the mass of such central object (the best known example is the controversial case of $\omega$ Cen, studied by \citealp{Noyola2008} and \citealp{vdMA2010}, with different conclusions). However, \citet{VesperiniTrenti2010} showed that these shallow photometric cusps are not decisive signatures of the presence of an IMBH, and that they might be related to other dynamical processes; moreover, the authors emphasize the fact that the typical accuracy in the data may be insufficient to characterize the slope of the profile as desired.

As shown in the previous sections, the presence of radially-biased pressure anisotropy (which occurs in the context of the family of $f^{(\nu)}$ models or of the Michie-King models, see \citealp{Michie1963}) can also produce a relatively rapid decline in the central part of the velocity dispersion profile. Therefore, here we are reiterating a point already noted in the literature, that the presence of a central IMBH should not be considered as the only physical explanation of the existence of a central kinematical peak\footnote{Of course, $f^{(\nu)}$ models are able to reproduce a slope in both the photometric and kinematic profile, only when high values of the concentration parameter are considered.}. Unfortunately, our independent conclusion only confirms that the interpretation of this interesting kinematical feature is more model dependent than desired. We recall that $f^{(\nu)}$ models are characterized by a ``realistic'' anisotropy profile (see Fig.~6 in \citealp{TrentiBertin2005_1}): the central regions are more isotropic than the outer ones in velocity space, because the models represent a scenario in which violent relaxation has acted more efficiently in the center. This differential kinematical feature is not always present in dynamical models based on the Jeans approach, which, to obtain a fast decline in the velocity dispersion profile in the absence of a central IMBH, usually requires very high values of the anisotropy parameter, even in the central regions of the cluster (see, for example, Sect~4.2 in \citealp{Noyola2008} or Sect~5.3 in \citealp{Lutz2011}).

\subsection{Radial orbit instability}
\label{Subsect.4.5}

Systems in which there is a great amount of radial kinetic energy with respect to tangential kinetic energy are subject to the radial orbit instability.

\citet{PS1981instab} (see also \citealp{FP1984instab}) introduced a parameter defined as:
\begin{equation}
\kappa = \frac{2 K_{\mathrm{r}}}{K_{\mathrm{T}}}
\label{instab1}
\end{equation}
where $K_{\mathrm{r}}$ and $K_{\mathrm{T}}$ represent the radial and the tangential component of the total kinetic energy, respectively, and argued that when:
\begin{equation}
\kappa > 1.7 \pm 0.25
\label{instab2}
\end{equation}
radial orbit instability occurs. Actually, different families of models generally have different threshold values for the instability. A detailed discussion of the onset of the radial orbit instability for the family of $f^{(\nu)}$ models can be found in \citet{TrentiBertin2005_1}, \citet{TrentiBertin2005_2}, and \citet{TrentiBertin2006}, where the validity of the criterion expressed in Eq. (\ref{instab2}) is discussed.

\begin{table}
\caption[]{Global anisotropy parameter for the best-fit $f^{(\nu)}$ models.}
\label{table8}
\centering
\begin{tabular}{cccccccc}
\hline\hline
\multicolumn{2}{c}{Relaxed} & & \multicolumn{2}{c}{Intermediate} & & \multicolumn{2}{c}{Partially Relaxed} \\
\cline{1-2} \cline{4-5} \cline{7-8}
NGC & $\kappa$ & & NGC & $\kappa$ & & NGC & $\kappa$ \\
\cline{1-2} \cline{4-5} \cline{7-8}
362  & 1.353 & & 6218 & 1.835 & & 5139 & 1.754 \\
7078 & 1.301 & & 6254 & 2.307 & & 2419 & 1.771 \\
104  & 1.301 & & 6656 & 1.443 & &      &       \\
6121 & 1.320 & & 3201 & 1.810 & &      &       \\
6341 & 1.443 & & 6809 & 1.859 & &      &       \\
     &       & & 288  & 1.859 & &      &       \\
\hline
\end{tabular}
\end{table}

We calculated the values of the parameter $\kappa$ for the $f^{(\nu)}$ models that provide the best fit to the observed profiles of the globular clusters of the sample (see Table~\ref{table8}). It is interesting to note that globular clusters belonging to a given relaxation class tend to have similar values of the stability parameter ($\kappa \approx 1.3,1.8,1.7$ for the first, second, and third relaxation class, respectively; in this respect, NGC 6656 and NGC 6254 appear to exhibit an exceptional behavior). In other words, the relaxed class is found to be more isotropic by the $f^{(\nu)}$ diagnostics.

The majority of the globular clusters that have $\kappa \ga 1.7$ are likely to be in a condition of \textit{marginal} instability. The case of NGC 6254 does remain problematic.

We argue that the introduction of a truncation in phase space to the family of $f^{(\nu)}$ models might have a stabilizing effect, since such truncation will affect primarily the outer parts of a given configuration, which are dominated by radially-biased pressure anisotropy. Therefore the truncation is likely to reduce the global value of the radial component of the total kinetic energy.

This is one more reason to undertake the study of truncated $f^{(\nu)}$ models for application to globular clusters. We plan it for the immediate future.


\section{Conclusions}
\label{Sect.5}

In this paper we have performed a detailed combined photometric and kinematic study of a sample of Galactic globular clusters, representing systems under different relaxation conditions. For these objects, surface brightness and velocity dispersion profiles have been fitted by means of two different families of dynamical models, the truncated isotropic King models and the non-truncated, anisotropic $f^{(\nu)}$ models. The analysis has been carried out by following the same procedure used in the past to study the dynamics of elliptical galaxies. Each globular cluster is then associated with two best-fit models. The main conclusions can be summarized as follows:
\begin{itemize}
\item The expected trend, that King models should perform better for more relaxed globular clusters, has been checked to exist but it is not as sharp as anticipated. The two clusters (NGC 104 and NGC 6341) for which the global fit by King models is most convincing indeed belong to the class of relaxed objects. King models tend to offer a good representation of the observed photometric profiles (as is commonly reported), regardless of the relaxation condition of the system (but a statement of this kind should also be supported by the relevant statistical indicators; see Table~\ref{table_stat}). However, the quality of the fits by King models to the kinematic profiles remains to be proved, even for relaxed clusters, because of the few data-points and the large error bars in the observed profiles. Three clusters for which the King models appear to be inadequate do not actually come as a surprise: NGC 2419 is the least relaxed cluster of the sample and NGC 362 and NGC 7078 are suspected to be post-core-collapse clusters.
\item The second expected trend, that less relaxed clusters might exhibit the characteristic signature of incomplete violent relaxation, is also partly present but is not as sharp as might have been hoped for. Some cases indeed point to a significant role of radially-biased pressure anisotropy. The least relaxed cluster, NGC 2419, is well described by the $f^{(\nu)}$ models. For the second least relaxed cluster, NGC 5139, the central shallow cusp in the velocity dispersion profile appears to be well captured by the $f^{(\nu)}$ models. A marginal indication in favor of the $f^{(\nu)}$ models also comes from inspection of the inner kinematic profiles of NGC 6218 and NGC 6254, although these two clusters are not among the least relaxed objects. In contrast, King models and other isotropic models (such as the spherical Wilson models) have difficulty in matching significant velocity gradients inside the half-light radius. Therefore, the partial success of the $f^{(\nu)}$ models suggests that for some globular clusters radially-biased pressure anisotropy may be important. This property could be examined further by means of better spatially-resolved kinematic data in the inner regions (i.e., at radii out to approximately the half-light radius). This result is in line with the conclusions of recent papers: in particular, see \citealp{Ibata2011} and references therein, based on the application of King-Michie models.
\item In some clusters, regardless of the relaxation condition, some qualitative characteristics of the observed profiles are missed by both families of models considered in this paper. It may be that part of these cases would be resolved by a study in terms of \textit{truncated} $f^{(\nu)}$ models. But it may also be that other ingredients ignored in the paper, such as rotation (solid-body or differential), play a role.
\item The paper demonstrates that the values of some structural parameters, such as the total mass and the half-mass radius, can be significantly model-dependent. In view of the results listed in the previous items, this is a clear warning against an indiscriminate use of structural parameters for globular clusters based on only one family of models (the spherical King models).
\item In general, the kinematic fits are crucial to assess if a model is actually suited to describe a given globular cluster. The main issue in testing dynamical models on globular clusters is therefore the general lack of good kinematic data: the data are available for a small fraction of the population of Galactic globular clusters and generally made of a small number of data-points, not well distributed in radius. Surprisingly, the kinematic profile is often not well sampled as desired inside the half-mass radius. As discussed in the paper this is a key region for confronting the performance of different dynamical models. In addition, accurate data in the outermost parts, close to the truncation radius,would touch on other important issues, such as the role of tides. Only after the acquisition of good kinematical profiles will it be possible to address properly the issue of dark matter in globular clusters.
\end{itemize}

\begin{acknowledgements}
We are grateful to D.C. Heggie for many constructive remarks that have helped improve the quality of the paper. We would like to thank E. Vesperini for a critical reading of the manuscript and valuable suggestions, M. Bellazzini, P. Bianchini, M. Lombardi, and C. Nipoti for useful comments and conversations, E. Noyola for providing us with the data for $\omega$ Cen, and S. E. Arena and M. Trenti for the routines to compute the profiles of the $f^{(\nu)}$ family of models. This work was partially supported by the Italian MIUR.
\end{acknowledgements}



\bibliographystyle{aa} 
\bibliography{biblio} 




\appendix


\section{The dynamical models}
\label{App.A}

\subsection{Spherical isotropic King models}
\label{App.A.1}

The \citet{King1966} one-component, spherical, isotropic models are based on the following distribution function:

\begin{equation}
f_{\mathrm{K}} = \bigg\{
\begin{array}{ll}
A\left[ \exp(-aE) - \exp(-aE_{\mathrm{0}}) \right] & E \leq E_{\mathrm{0}} \\
0 & E > E_{\mathrm{0}} \ \ , \\
\end{array}
\label{kingdf}
\end{equation}

\noindent where $A$, $a$, $E_{\mathrm{0}}$ are positive constants, defining two scales and one dimensionless parameter, and $E$ represents the specific energy $E =v^2/2 + \Phi(r)$, where $\Phi(r)$ is the mean-field gravitational potential, to be determined from the Poisson equation. The quantity $E_{\mathrm{0}}$ is a threshold energy, above which the stars are considered unbound; it can be translated into a truncation radius, $r_{\mathrm{tr}}$, for the system.

We recall that the radial scale $r_{\mathrm{0}}$ can be defined as

\begin{equation}
r_{\mathrm{0}} = \left( \frac{9}{4 \pi G a \rho_{\mathrm{0}}} \right)^{1/2} \ ,
\label{r0}
\end{equation}

\noindent where $\rho_{\mathrm{0}}$ is the central mass density. The other dimensional scale is the total mass of the cluster or, as an alternative, its central velocity dispersion.

The dimensionless concentration parameter is given by the central dimensionless potential $\Psi = a[\Phi(r_{\mathrm{tr}}) - \Phi(0)]$ or, alternatively, by the index

\begin{equation}
c = \log\left(\frac{r_{\mathrm{tr}}}{r_{\mathrm{0}}} \right) \ ,
\label{c}
\end{equation}

\noindent because these two quantities are in a one-to-one relation.

The quantity listed in the Harris catalog as concentration parameter, in this paper indicated with $C$, is defined using the standard core radius $R_{\mathrm{c}}$ in place of $r_{\mathrm{0}}$ in Eq. (\ref{c}). The values of $C$ for a model identified by $\Psi \ga 4$ are slightly larger than the corresponding values of $c$; in fact, for these models $0.8 \la R_{\mathrm{c}}/r_{\mathrm{0}} \la 1$.

\subsection{Anisotropic $f^{(\nu)}$ models}
\label{App.A.2}

Several families of dynamical models have been developed to represent the final state of numerical simulations of the violent relaxation process thought to be associated with the formation of bright elliptical galaxies via collisionless collapse (for a review, see \citealp{BertinStiavelli1993}). These models show a characteristic anisotropy profile, with an inner isotropic core and an outer envelope that becomes dominated by radially-biased anisotropic pressure. They provide a good representation of the photometric and kinematic properties of elliptical galaxies. Here we will refer to the family of spherical, anisotropic, non-truncated $f^{(\nu)}$ models (which have been revisited recently in detail by \citealp{BertinTrenti2003}).

The distribution function that defines these models depends on specific energy $E$ and angular momentum $J$:

\begin{equation}
f^{(\nu)} = \bigg\{
\begin{array}{ll}
A \exp \left[ -aE - d\left( \frac{J^2}{|E|^{3/2}} \right)^{\nu/2} \right] & E
\leq 0 \\
0 & E > 0 \ \ , \\
\end{array}
\label{fnu}
\end{equation}

\noindent  where $A$, $a$, $d$, and $\nu$ are positive constants, defining two scales and two dimensionless parameters. For applications, as described by \citet{BertinTrenti2003}, the dimensionless parameter $\nu$ can be fixed at $\nu = 1$. Therefore, similarly to the King models, after integration of the relevant Poisson equation the $f^{(\nu)}$ models are a one-parameter family of models, parametrized by their central concentration, which can be expressed by the central dimensionless potential $\Psi = -a\Phi(0)$.

The physical scales can be expressed as:

\begin{equation}
\begin{array}{l}
r_{\mathrm{scale}} = d^{-1/\nu}a^{-1/4} \\
M_{\mathrm{scale}} = d^{-3/\nu}a^{-9/4} \ . \\
\end{array}
\label{scalesfnu}
\end{equation}

By definition, these models are non-truncated and, because of this, are likely to be less suited to describe the outer parts of globular clusters. A study of globular clusters based on truncated $f^{(\nu)}$ models is postponed to a separate investigation.

We recall that the surface brightness profiles for concentrated models ($\Psi \ga 7$) are very close to de Vaucouleurs profile, while for low values of $\Psi$ the models exhibit a sizeable core.


\section{Fitting procedure}
\label{App.B}

In the following we describe the procedure that we have adopted to perform the statistical analysis of the data. We basically follow \citet{BSS88}.

\subsection{Photometric fit}
\label{App.B.1}

For each family of models considered in this paper, the photometric fit determines which equilibrium model has the projected mass distribution that best reproduces the surface brightness profile, under the assumption that:

\begin{equation}
\Sigma(R) = \frac{M}{L} \lambda(R) \ ,
\label{moverl}
\end{equation}

\noindent where $\Sigma(R)$ is the model surface mass density, $\lambda(R)$ the model surface luminosity density, and the mass-to-light ratio $M/L$ is considered to be constant in the cluster. From the photometric data, we perform a fit that allows us to determine three parameters for each model:

\begin{itemize}
\item $\Psi$, the concentration parameter, which determines the shape of the surface brightness and velocity dispersion profiles;
\item $r_{\mathrm{s}}$, the scale radius, that is $r_{\mathrm{0}}$ for the King models and $r_{\mathrm{scale}}$ for the $f^{(\nu)}$ models;
\item $\mu_{\mathrm{0}}$, the central surface brightness.
\end{itemize}

We might determine the values of these parameters by minimizing the quantity:

\begin{equation}
\sum_{i = 1}^{N_{\mathrm{p}}}\left[ \frac{m(R_i) - S(R_i/r_{\mathrm{s}}) +
\mu_0}{\delta m_i} \right]^2 \ ,
\label{chimag}
\end{equation}

\noindent where $S$ is the surface density expressed in magnitudes, but we found that it is more convenient to minimize:

\begin{equation}
\chi^2_{\mathrm{p}} = \sum_{i = 1}^{N_{\mathrm{p}}}\left[ \frac{l(R_i) -
k\hat{\Sigma}(R_i/r_{\mathrm{s}})}{\delta l_i} \right]^2 \ ,
\label{chi}
\end{equation}

\noindent because the chi-squared function turns out to be more stable, as pointed out by \citet{MLvdM2005}. The connection between Eqs. (\ref{chimag}) and (\ref{chi}) is easily made, by taking into account the following relations:

\begin{equation}
\begin{array}{l}
m(R_i) - m_{0} = -2.5 \log[l(R_i)] \\
S(R_i/r_{\mathrm{s}}) = -2.5 \log[\hat{\Sigma}(R_i/r_{\mathrm{s}})] \\
\mu_0 - m_{0} = 2.5 \log k \ ,
\end{array}
\label{lummag}
\end{equation}

\noindent where the zero-point $m_{0} = 26.422$ allows us to express the observed surface brightness $m(R_i)$ in solar units; the observed luminosities $l(R_i)$ are expressed in solar units and $\delta l_i$ are calculated from $\delta m_i$ and represent the errors on luminosities. The quantity $\hat{\Sigma}(R_i/r_{\mathrm{s}})$ is the surface density, normalized to its central value; from this point of view, the parameter $k$ is related to the mass-to-light ratio.

Even if we calculated the parameters from the luminosities, we report the results in terms of magnitudes, because in this way a comparison with the data is more natural. To calculate the surface brightness profile $\mu(R)$ that best reproduces the data, we choose the model identified by $\Psi$, we calculate the normalized projected mass-density profile, and then we rescale it radially with $r_{\mathrm{s}}$ and vertically with $\mu_{\mathrm{0}}$.

Note that we have only two independent fit parameters, because $k = k(\Psi, r_{\mathrm{s}})$. Hence, the value of the reduced $\chi^2_{\mathrm{p}}$ is:

\begin{equation}
\widetilde{\chi}^2_{\mathrm{p}} = \frac{\chi^2_{\mathrm{p}}}{N_{\mathrm{p}}-2} \ .
\label{chired}
\end{equation}

\noindent To have a quantitative estimate of the quality of the fit, it is convenient to calculate also the mean and maximum residuals:

\begin{equation}
\begin{array}{l}
\langle \Delta \mu \rangle = \frac{1}{N_{\mathrm{p}}^{1/2}}\left\lbrace \sum_{i = 1}^{N_{\mathrm{p}}}[m(R_i) - \mu(R_i)]^2 \right\rbrace ^{1/2} \\
(\Delta \mu)_{\mathrm{max}} = \max_{i = 1, \ldots, N_{\mathrm{p}}} |m(R_i) - \mu(R_i)| \ . \\
\end{array}
\label{residuals}
\end{equation}

\subsection{Kinematic fit}
\label{App.B.2}

The parameters identified by the photometric fit determine a model, characterized by the parameter $\Psi$, the profiles of which are rescaled with $r_{\mathrm{s}}$. At this point, we perform a fit to the kinematic data to find the central line-of-sight velocity dispersion, that is the velocity scale needed to rescale vertically the normalized projected velocity dispersion profile calculated from the dynamical models. We calculate the value of this scale, $V$, as the one that minimizes:

\begin{equation}
\chi^2_{\mathrm{k}} = \sum_{i = 1}^{N_{\mathrm{k}}} \left[ \frac{\sigma(R_i) - V
\sigma_{\mathrm{P}}(R_i)}{\delta \sigma_i} \right]^2 \ ,
\label{kinchi}
\end{equation}

\noindent where $\sigma_{\mathrm{P}}(R_i)$ is the projected velocity dispersion calculated from the model, dependent on the rescaled radial coordinate. In this case, the values of the reduced $\chi^2$ and of the residuals can be found from the following expressions:

\begin{equation}
\begin{array}{l}
\widetilde{\chi_{\mathrm{k}}}^2 = \frac{\chi_{\mathrm{k}}^2}{N_{\mathrm{k}}-1}
\\
\langle \Delta \sigma\rangle = \frac{1}{N_{\mathrm{k}}^{1/2}}\left\lbrace \sum_{i = 1}^{N_{\mathrm{k}}}[\sigma(R_i) - V \sigma_{\mathrm{P}}(R_i)]^2
\right\rbrace ^{1/2} \\
(\Delta \sigma)_{\mathrm{max}} = \max_{i = 1, \ldots, N_{\mathrm{k}}} |\sigma(R_i) - V \sigma_{\mathrm{P}}(R_i)| \ . \\
\end{array}
\label{kinres}
\end{equation}

One might argue that a more sensible way to perform the fit would be by means of a single fit procedure, using the combined $\chi_{\mathrm{tot}}^2=\chi_{\mathrm{p}}^2+\chi_{\mathrm{k}}^2$, defined as the sum of the photometric and of the kinematic contribution. We did perform tests of this different procedure and found that the results are equivalent to those obtained by performing the photometric fit first and then the kinematic fit at fixed $\Psi$ and $r_{\mathrm{s}}$. This confirms the qualitative expectation that the kinematical data, being less numerous and less accurate with respect to the photometric ones, have little weight in determining  $\Psi$ and $r_{\mathrm{s}}$ and are only needed to determine the scale $V$, that is, the relevant mass-to-light ratio.

\begin{figure*}[t]
\centering
\includegraphics[height=17cm, angle=270]{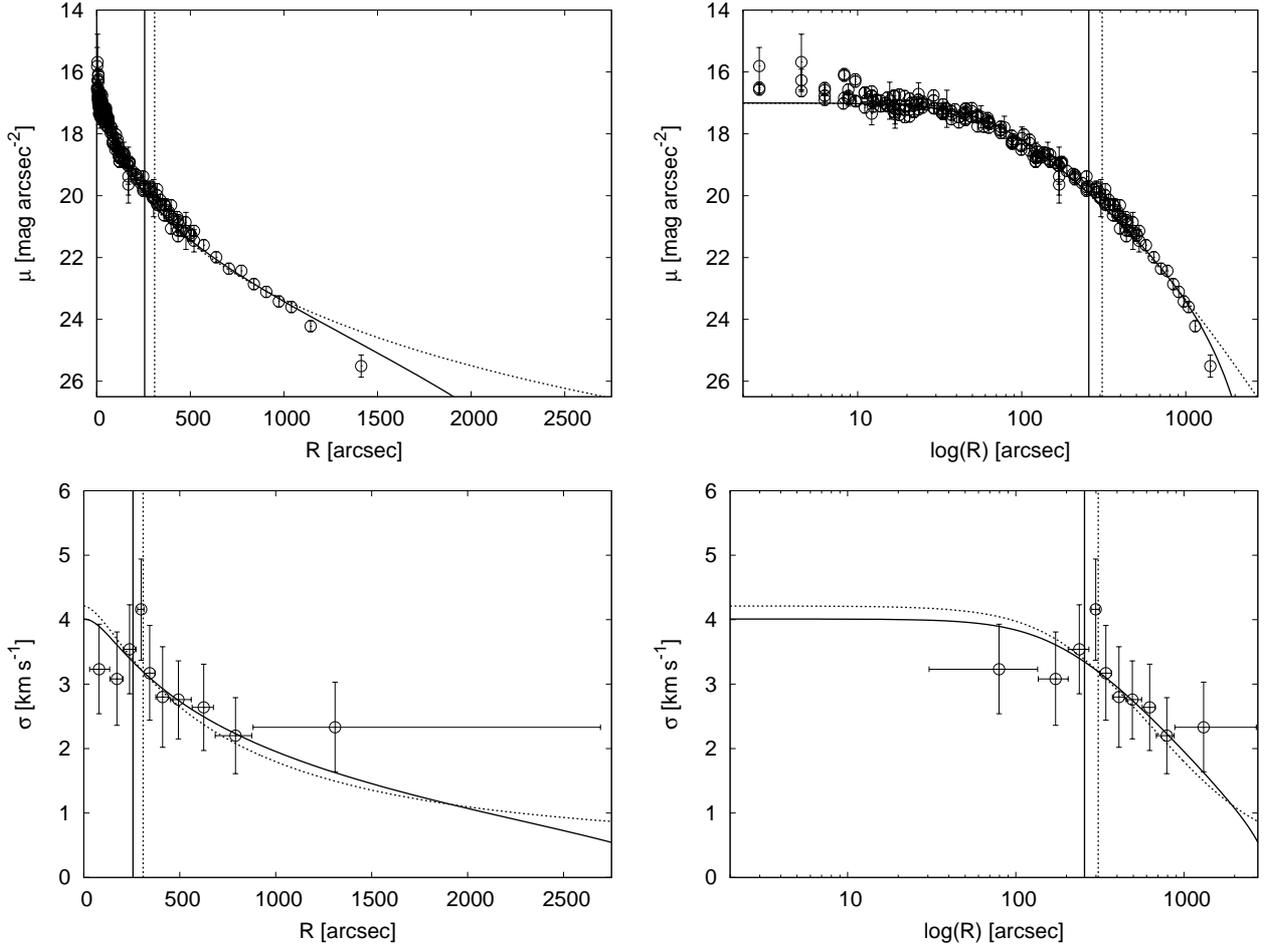}
\caption{Linear vs logarithmic representation for the photometric (upper panels) and kinematic (lower panels) profiles of NGC 6121. On the left, profiles are plotted with a linear radial scale, on the right with a logarithmic scale.}
\label{figureB2}
\end{figure*}

Finally, we addressed the issue of whether a different representation of the results of the best-fit models would be more suitable than the one used in Figs.~\ref{figure1}-\ref{figure3}. Figure~\ref{figureB2} compares the appearance of a fit for a given object using a linear and a logarithmic representation of the radial scale. Based on this and other tests, we preferred to adopt a mixed representation, as used in the main text of the paper.

\subsection{Goodness of the photometric and kinematic fits}
\label{App.B.Stat}

To measure the goodness of the photometric and kinematic fits, we referred to the confidence intervals on the $\chi^2$-distribution. The probability density function of the $\chi^2$ distribution is defined as:

\begin{equation}
f(x;n) = \frac{e^{(-x/2)} x^{(n/2-1)}}{2^{n/2}\Gamma(n/2)}
\label{stat.1}
\end{equation}

\noindent when $x \geq 0$, and zero otherwise. The parameter $n$ corresponds to the number of degrees of freedom and $\Gamma$ denotes the gamma function (e.g., see \citealp{AbramowitzStegun}, Sect.~26.4).

The two-sided confidence interval relative to a confidence level $\alpha$ on the $\chi^2$ probability density function is defined as the interval $[\chi^2_{\mathrm{inf}},\chi^2_{\mathrm{sup}}]$ such that
\begin{equation}
\alpha = \int_{\chi^2_{\mathrm{inf}}}^{\chi^2_{\mathrm{sup}}} f(x;n)dx \ ;
\end{equation}
\noindent and that:
\begin{equation}
1-\frac{\alpha}{2} =  \int_{0}^{\chi^2_{\mathrm{inf}}} f(x;n) \, dx =  \int_{\chi^2_{\mathrm{sup}}}^{\infty} f(x;n) \, dx \ .
\end{equation}

\subsection{Errors on the best-fit parameters}
\label{App.B.3}

\begin{figure*}
\centering
\includegraphics[height=17cm, angle=270]{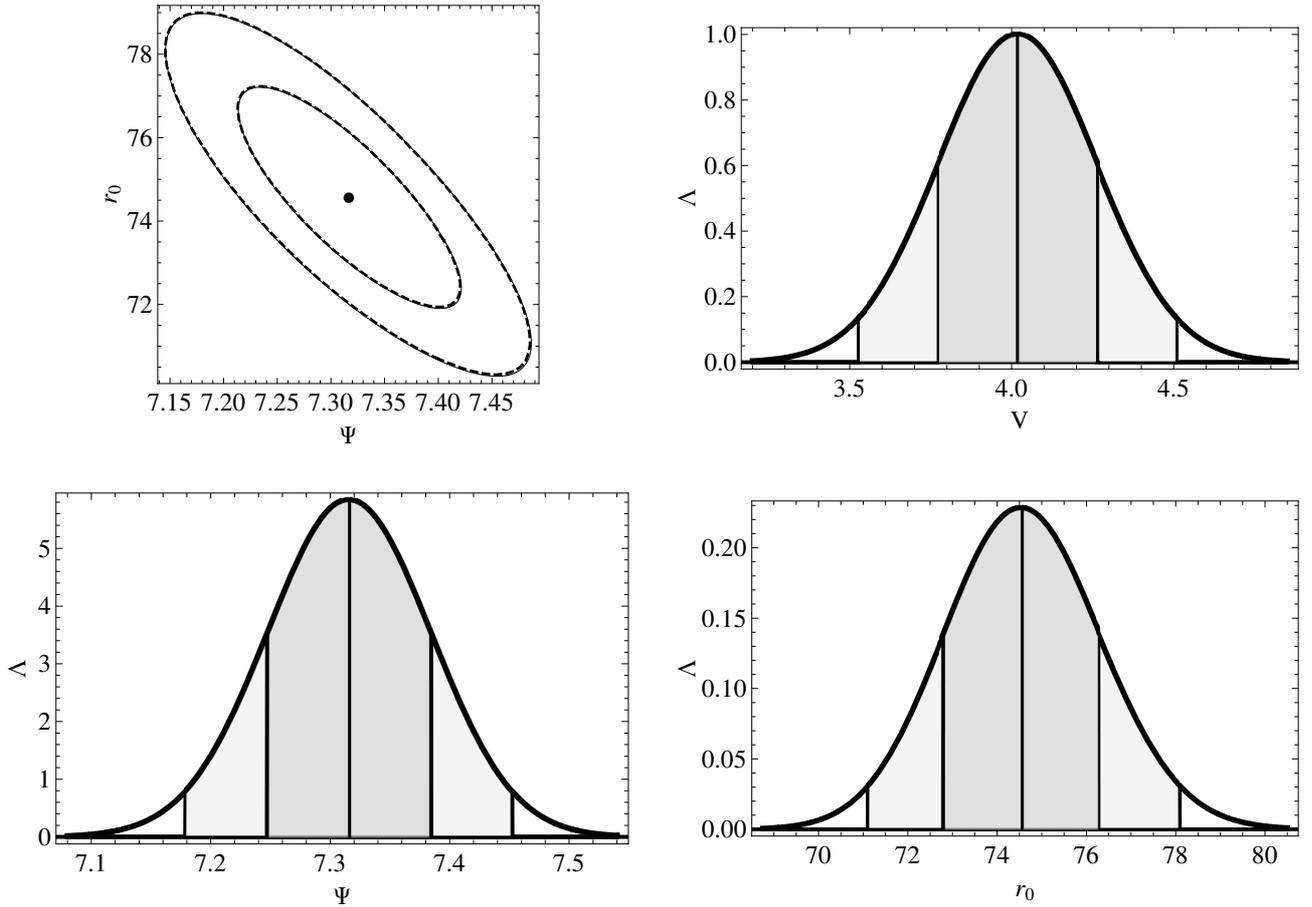}
\caption{Confidence regions and confidence intervals on NGC 6121 King model parameters. The top left figure shows the overlap of the confidence regions, corresponding to confidence levels of $68.3\%$ and $95.4\%$ (solid lines), and the curves of constant $\chi^2$ (dashed lines); the black dot marks the position of the maximum of the likelihood (that is the minimum of $\chi^2$). The top right figure shows the confidence intervals for the likelihood depending on $V$; the bottom figures show the confidence intervals on the parameter $\Psi$ and $r_{\mathrm{0}}$; the dark grey area corresponds to the confidence level of $68.3\%$, the light grey area to the confidence level of $95.4\%$; a solid vertical line marks the position of the maximum of the likelihood (that is the minimum of $\chi^2$).}
\label{figureB1}
\end{figure*}

For the methods described below, we refer mainly to \citet{NR}. In the case of the photometric fit, we used the following procedure to calculate the formal errors on the parameters. First we calculated the Hessian matrix:

\begin{equation}
H_{ij} = \frac{\partial^2\chi^2}{\partial x_i\partial x_j} \ ,
\label{hess}
\end{equation}

\noindent where $i = 1,2,3$ and $x_{\mathrm{1}} = \Psi$, $x_{\mathrm{2}} = r_{\mathrm{s}}$ and $x_{\mathrm{3}} = \mu_{\mathrm{0}}$; then we calculated the covariance matrix:

\begin{equation}
[E] = 2[H]^{-1} \ .
\label{cov}
\end{equation}

\noindent Finally, we obtained the errors: $\delta x_i = ({E_{ii}})^{1/2}$.

In the case of the kinematic fit, in which $\chi_{\mathrm{k}}^2$ depends analytically only on the parameter $V$, we can immediately calculate:

\begin{equation}
\delta V = \left( {\frac{2}{\partial^2\chi_{\mathrm{k}}^2/\partial V^2}} \right)^{1/2} \ .
\label{kinerr}
\end{equation}

In addition to errors, we identified the relevant confidence regions and intervals for the various parameters, after defining a likelihood function $\Lambda = e^{-\chi^2/2}$. First, we calculated confidence regions in the $(\Psi, r_{\mathrm{s}})$ plane, after marginalizing the likelihood function on $k$, in the following way:

\begin{equation}
\Lambda(\Psi, r_{\mathrm{s}}) = \int e^{-\chi^2/2} dk \ .
\label{lam}
\end{equation}

\noindent We identified the regions corresponding to confidence levels of $68.3\%$ and $95.4\%$. To check the results, we compared them with the regions defined by the curves of constant $\chi^2$ with the appropriate values \citep[see Sect.~15.6 in][]{NR}. For each globular cluster and for the families of models, we notice that the regions identified with these two methods overlap in a consistent way.

We also calculated the confidence intervals on the parameters $\Psi$, $r_{\mathrm{s}}$ and $V$ (for the last parameter, the calculation can be done immediately, because the kinematic likelihood depends only on $V$; for the other parameters, we have to marginalize once more the likelihood function), corresponding to confidence levels of $68.3\%$ and $95.4\%$. Comparing them with the errors calculated with the covariance matrix method, we found that the results are consistent.

In Fig.~\ref{figureB1} we show an example of the confidence regions and intervals calculated for the King best-fit model for the globular cluster NGC 6121. The top left panel shows the overlap of the confidence regions (solid lines) corresponding to confidence levels of $68.3\%$ and $95.4\%$, and the curves of constant $\chi^2$ (dashed lines). It is clear that there is a good agreement between the two relevant pairs of curves. The bottom panels show the confidence intervals calculated on the marginalized likelihood for the parameters $\Psi$ and $r_{\mathrm{0}}$, and the top right panel those for the parameter $V$; in these figures the confidence intervals are shown as shaded areas. By comparing the extent of the dark grey area with the values of the uncertainties on the corresponding parameters (see Table~\ref{table3}), we see that the different methods lead to consistent results.

\subsection{Derived parameters}
\label{App.B.4}

Once the best-fit parameters have been secured, we may calculate other quantities, which represent some important characteristics of globular clusters.

In particular, for each family of models we calculated the core radius $R_{\mathrm{c}}$, that is the radial position where the surface brightness equals half its central value, the half-mass radius $r_{\mathrm{M}}$, the total cluster mass $M$, the central mass density $\rho_{\mathrm{0}}$, the mass-to-light ratio $M/L$, the core relaxation time $\log{T_{\mathrm{c}}}$ and the half-mass relaxation time, $\log{T_{\mathrm{M}}}$. In addition, for King models we calculated the concentration parameter $c$ and the truncation radius $r_{\mathrm{tr}}$. In turn, for the $f^{(\nu)}$ models we calculated the anisotropy radius $r_{\alpha}$, defined as the radial position where $\alpha(r_{\alpha}) = 1$, with
\begin{equation}
\alpha = 2 - (\langle v_{\theta}^2\rangle + \langle v_{\phi}^2\rangle)/\langle v_r^2\rangle \ .
\label{alpha}
\end{equation}

The ratio of this radius to the half-mass radius measures how large is the radially-biased anisotropic part of the globular cluster: the smaller $r_{\alpha}$, the larger the anisotropic region of the cluster.

\end{document}